\documentclass[journal]{IEEEtran}

\usepackage{tikz}
\usepackage{amsmath}
\usepackage{amsfonts}
\usepackage{algorithm}
\usepackage{algorithmic}
\usepackage{graphicx}
\usepackage{textcomp}
\usepackage{xcolor}
\usepackage{array}
\usepackage{multicol, multirow}
\usepackage{makecell}
\usepackage{booktabs}
\usepackage[caption=false,font=normalsize,labelfont=sf,textfont=sf]{subfig}
\usepackage{cite}
\usepackage{stfloats}
\usepackage{url}
\usepackage{verbatim}
\usepackage{hyperref}
\newcommand{\secref}[1]{Sec.~\ref{#1}}

\newcommand{\figref}[1]{Fig.~\ref{#1}}
\newcommand{\tabref}[1]{Table~\ref{#1}}

\newcommand{\equref}[1]{Eq.~(\ref{#1})}

\newcommand{\subsecref}[1]{Subsec.~\ref{#1}}

\begin{document}
%-------------------------------------------------------------------------------

\title{Eunomia: A Multicontroller Domain Partitioning Framework \\ in Hierarchical Satellite Networks}

\author{Qi Zhang, Kun Qiu, Zhe Chen, Wenjun Zhu, Xiaofan Xu, Ping Du and Yue Gao% <-this % stops a space
\thanks{
Qi Zhang, Kun Qiu, Zhe Chen, Wenjun Zhu, and Yue Gao are with the School of Computer Science and Intelligent Networking and Computing Research Center, Fudan University, Shanghai 200433, China (e-mail: qizhang23@m.fudan.edu.cn, \{qkun, zhechen, wenjun, gao.yue\}@fudan.edu.cn). 
Xiaofan Xu and Ping Du are with the State Key Laboratory of Satellite Network and Shanghai Satellite Network Research Institute Co., Ltd., Shanghai 201210, China (e-mail: xuxf@chinasatnet.com.cn; pingdu@ustc.edu.cn). 
 The corresponding author is Kun Qiu, and the co-corresponding author is Xiaofan Xu.}% <-this % stops a space
} % end author

\markboth{IEEE Journal on Selected Areas in Communications}{Author \MakeLowercase{\textit{et al.}}: Eunomia}

\maketitle
%-------------------------------------------------------------------------------
\begin{abstract}
%-------------------------------------------------------------------------------
With the rise of mega-satellite constellations, the integration of hierarchical non-terrestrial and terrestrial networks has become a cornerstone of 6G coverage enhancements. In these hierarchical satellite networks, controllers manage satellite switches within their assigned domains. However, the high mobility of LEO satellites and field-of-view (FOV) constraints pose fundamental challenges to efficient domain partitioning. Centralized control approaches face scalability bottlenecks, while distributed architectures with onboard controllers often disregard FOV limitations, leading to excessive signaling overhead. LEO satellites outside a controller's FOV require an average of five additional hops, resulting in a 10.6-fold increase in response time.
To address these challenges, we propose Eunomia, a three-step domain-partitioning framework that leverages movement-aware FOV segmentation within a hybrid control plane combining ground stations and MEO satellites. Eunomia reduces control plane latency by constraining domains to FOV-aware regions and ensures single-hop signaling. It further balances traffic load through spectral clustering on a Control Overhead Relationship Graph and optimizes controller assignment via the Kuhn-Munkres algorithm.
We implement Eunomia on the Plotinus emulation platform with realistic constellation parameters. Experimental results demonstrate that Eunomia reduces request loss by up to 58.3\%, control overhead by up to 50.3\%, and algorithm execution time by 77.7\%, significantly outperforming current state-of-the-art solutions.
\end{abstract}


%-------------------------------------------------------------------------------
\section{Introduction}
%-------------------------------------------------------------------------------

With the advent of 6G and NextG visions for ubiquitous connectivity, the fusion of terrestrial and Non-Terrestrial Networks (NTN) has emerged as a key enabler of global coverage and service resilience~\cite{hanEvolution6GSatellite2023}. In this context, the rapid rollout of mega-satellite constellations is revolutionizing long-haul and remote communications~\cite{luo2024hurry}. By 2024, SpaceX's Starlink had secured over 4 million subscribers~\cite{degaudenziSatelliteNetworksPresent2025}, underscoring the commercial viability of large-scale Low Earth Orbit (LEO) deployments. Yet, as planned constellations grow to tens of thousands of satellites~\cite{xieLEOMegaconstellations6G2021}, persistent challenges in network control and domain management threaten to undermine operational efficiency and end-to-end quality of service (QoS).

To address the increasing complexity and scale of satellite networks, hierarchical satellite networks have gained traction as a promising architectural solution for distributed control. A central problem in hierarchical satellite networks is the controller deployment and domain partitioning problem: determining controller assignments for satellite subsets. This decision directly affects system performance, control overhead, and QoS. Real-world deployments highlight the severity of this challenge. For instance, Starlink employs a centralized control scheme, issuing global updates every 15 seconds~\cite{tanveerMakingSenseConstellations2023, panMeasuringSatelliteLinks2024, liu2024efficient}, covering satellite selection, routing, and flow table updates. While conceptually straightforward, this approach incurs significant control latency and periodic overhead surges that strain the network during peak load. Additionally, inter-satellite link (ISL) instability further degrades performance; for every 1\% decrease in ISL stability, control message retransmissions increase by 13-24\%~\cite{yangInterruptionToleranceStrategy2023, choCrossLayerOptimization2019}, leading to excessive delays and reduced responsiveness.

Numerous domain partitioning strategies have been proposed to address these scalability limitations, yet each suffers from critical limitations. Centralized ground station (GS) control offers global optimization but struggles with scalability and incomplete LEO coverage as constellation sizes increase. 
Distributed LEO control~\cite{liOptimizedControllerProvisioning2023} improves scalability by delegating control to LEO nodes, but imposes substantial computational and power burdens on resource-constrained satellites. 
Hierarchical Middle Earth Orbit (MEO) or Geostationary Earth Orbit (GEO) control~\cite{chenHierarchicalDomainBasedMulticontroller2022} introduces intermediate control layers to reduce load, but the resulting propagation delays—up to 260ms for GEO links—are unacceptable for real-time applications. Alternatives such as stateless routing~\cite{liuStatelessDesignSatellite2024} and MEO backbone architectures offer partial solutions but lack sufficient dynamic traffic management and fail to meet the distributed control demands of commercial satellite networks.

Beyond these limitations, a critical and largely overlooked issue lies in the handling of satellite mobility under field-of-view (FOV) constraints. Existing MEO-LEO hybrid control approaches~\cite{xuDistributedMultilayerHierarchical2024} model FOVs as geometric visibility regions but ignore the satellites' predictable directional movement patterns, such as northbound and southbound trajectories. This mobility-agnostic design leads to frequent fragmentation of the control topology, unnecessary reconfigurations, and unstable domain boundaries, ultimately impairing network efficiency and reliability.

To address these challenges, we propose Eunomia, a novel three-step domain partitioning framework that leverages a unified hybrid control plane integrating both ground stations and MEO satellites. Eunomia introduces three key innovations: Movement-aware FOV partitioning, which accounts for satellite directional mobility to reduce domain boundary instability; Control Overhead Relationship Graph (CORG) modeling, which captures the complex interdependencies between control operations and evolving network topologies; and a progressive three-step optimization process which balances FOV constraints, control overhead, and load distribution using spectral clustering and the Kuhn-Munkres algorithm for optimal matching.

In contrast to conventional approaches that treat FOV as a static geometric constraint, Eunomia exploits the deterministic orbital mechanics of satellites to build stable, mobility-aware control domains. Our hybrid GS-MEO control plane ensures full LEO coverage with minimal controller count, while the progressive optimization algorithm guarantees both computational efficiency and high solution quality.

We implement Eunomia in a satellite network emulation platform based on real constellation parameters from Starlink, OneWeb, and other operational systems. Experimental results show that Eunomia outperforms existing approaches, achieving a 19.6\%-58.3\% reduction in request loss rates, a 19.95\%-50.3\% reduction in control overhead, and up to a 77.7\% reduction in algorithm execution time, thus demonstrating its practical viability for large-scale satellite network deployments.

This paper makes the following three primary contributions to the field of satellite network control:
\begin{enumerate}
    \item Problem Formulation: we present the first comprehensive mathematical formulation of control overhead in FOV-constrained satellite networks, framing domain partitioning as an NP-hard optimization problem that incorporates directional mobility and control dependencies.
    \item Mobility-Aware Domain Partitioning: we propose Eunomia, a novel three-step framework that integrates satellite mobility patterns into the domain partitioning process, achieving stable, balanced, and low-overhead control domains through progressive optimization.
    \item System Implementation and Evaluation: we develop a full-featured implementation of Eunomia, including detailed control protocols, and validate its performance through extensive emulation experiments based on realistic satellite network scenarios.
\end{enumerate}

The rest of this paper is organized as follows. \secref{sec_bac} introduces the background and challenges of hierarchical satellite networks. \secref{sec_ove} introduces the overall design of Eunomia. \secref{sec_mod}, \ref{sec_alg} and \ref{sec_sys_imp} detail the control overhead model, the domain partitioning algorithm and the system implementation necessary for the deployment algorithm, respectively. \secref{sec_eva} presents experimental evaluation results. Finally, \secref{sec_con} concludes the paper. A summary of key notations used throughout the paper is provided in \tabref{tab_notation}.

\begin{table}[t]
  \caption{Key Notation Summary}
  \label{tab_notation}
  \centering
  \begin{tabular}{cl}
    \toprule
    \textbf{Notation} & \textbf{Description} \\
    \midrule
    $V$ & Set of LEO satellites \\
    $K$ & Set of controllers (GS and MEO satellites) \\
    $D$ & Set of control domains \\
    $T$ & Set of time slots \\
    $W_{\text{CTL}}(t)$ & Total control overhead at time $t$ \\
    $W_{\text{FLOW}}(t)$ & Flow table update overhead \\
    $W_{\text{SYNC}}(t)$ & Domain synchronization overhead \\
    $W_{\text{MIG}}(t)$ & Controller migration overhead \\
    $W_{\text{CPT}}(t)$ & Path computation overhead \\
    $x_{i,j}^t$ & Binary variable: satellites $i$, $j$ in same domain \\
    $y_{d,k}^t$ & Binary variable: domain $d$ managed by controller $k$ \\
    $\lambda_{i,j}^t$ & Flow arrival rate between satellites $i$ and $j$ \\
    $M_{\text{fl}}$ & Flow table message size (36 bytes) \\
    $M_{\text{sync}}$ & Synchronization message size (24 bytes) \\
    $h_{i,j}$ & Number of hops from satellite $i$ to controller \\
    $f_{\text{sync}}$ & Synchronization frequency \\
    $f_{\text{mig}}^d$ & Migration frequency \\
    $\xi_{i,j}$ & CORG edge weight between satellites $i$ and $j$ \\
    \bottomrule
  \end{tabular}
\end{table}

%-------------------------------------------------------------------------------
\section{Background} \label{sec_bac}
%-------------------------------------------------------------------------------

\subsection{Hierarchical Satellite Network}

The deployment of large-scale LEO constellations has accelerated rapidly in recent years. By 2024, SpaceX's Starlink had launched over 7,500 satellites, Amazon's Project Kuiper plans to deploy 3,236 satellites~\cite{delportilloTechnicalComparisonThree2019}, and Chinese ``GW'' constellation aims to operate a network of 13,000 LEO satellites~\cite{xuSelforganizingControlMega2022}. In parallel, MEO constellations—such as OneWeb, supported by O3b—have validated the feasibility of hierarchical constellation integration~\cite{osoroTechnoeconomicFrameworkSatellite2021}. 
These advancements have led to the development of hierarchical satellite networks, which integrate MEO and LEO layers (sometimes incorporating GEO layers) to achieve global coverage, enhance network scalability, and optimize resource utilization~\cite{chenSurveyResourceManagement2024, wang2025stabilizing, peng2025dora}.

In satellite networks, the control plane operates with a clear separation between controllers and switches. Controllers (deployed on GS and MEO satellites) maintain global network views and make routing decisions, while LEO satellites function as transparent forwarding switches that execute controller instructions through flow tables. Each controller manages a domain of LEO switches, periodically synchronizing network state and updating flow tables in response to traffic demands.

\subsection{Limitations of Existing Domain Partitioning Approaches}

However, existing hierarchical domain partitioning approaches suffer from two fundamental problems that limit their practical deployment in large-scale satellite networks.

\textbf{Problem 2.1: Overly Idealized Control Models.} Existing control models for satellite networks fail to adequately capture control overhead at the packet level, especially regarding the effects of LEO inter-satellite link instability.~\cite{liOptimizedControllerProvisioning2023, tanveerMakingSenseConstellations2023}. Empirical studies reveal that centralized approaches like Starlink's 15-second global routing updates~\cite{tanveerMakingSenseConstellations2023, panMeasuringSatelliteLinks2024} and distributed LEO-based control~\cite{liStableHierarchicalRouting2024} fail to account for realistic control plane challenges. Specifically, satellite processors handle OpenFlow operations in under 100$\mu$s while propagation delays range from 1 to 40ms~\cite{georgeOnboardProcessingHybrid2018}, yet existing models ignore the amplified impact of ISL instability on control message transmission. Studies show that processing delays typically account for less than 5\% of total control overhead~\cite{chenHierarchicalDomainBasedMulticontroller2022}, indicating that current models significantly underestimate the impact of network-layer factors on control plane performance.

\textbf{Problem 2.2: Inadequate FOV and Mobility Considerations.} Current approaches either completely ignore FOV constraints~\cite{maImplementationCentralizedControllers2022} or consider them while overlooking satellite movement direction patterns~\cite{xuDistributedMultilayerHierarchical2024}. This leads to frequent control domain topology updates and reduced control plane reliability. Quantitative analysis shows that LEO satellites outside controller FOV require an average of 5 additional hops compared to those within FOV, resulting in 10.6$\times$ longer response times~\cite{chenMobilityLoadadaptiveController2021}. Furthermore, when considering only MEO-to-LEO FOV constraints ($\pm 40^{\circ}$ elevation threshold~\cite{plastikovHighgainMultibeamBifocal2016}) while ignoring orbital plane movement directions, control topology updates occur 2.3$\times$ more frequently than movement-aware approaches~\cite{huangEfficientDifferentiatedRouting2024}. These frequent updates directly impact control signaling timeliness and network fault detection capabilities.

\subsection{Key Technical Concepts}

\begin{figure}[t]
  \centering
  \includegraphics[width= 2.8in]{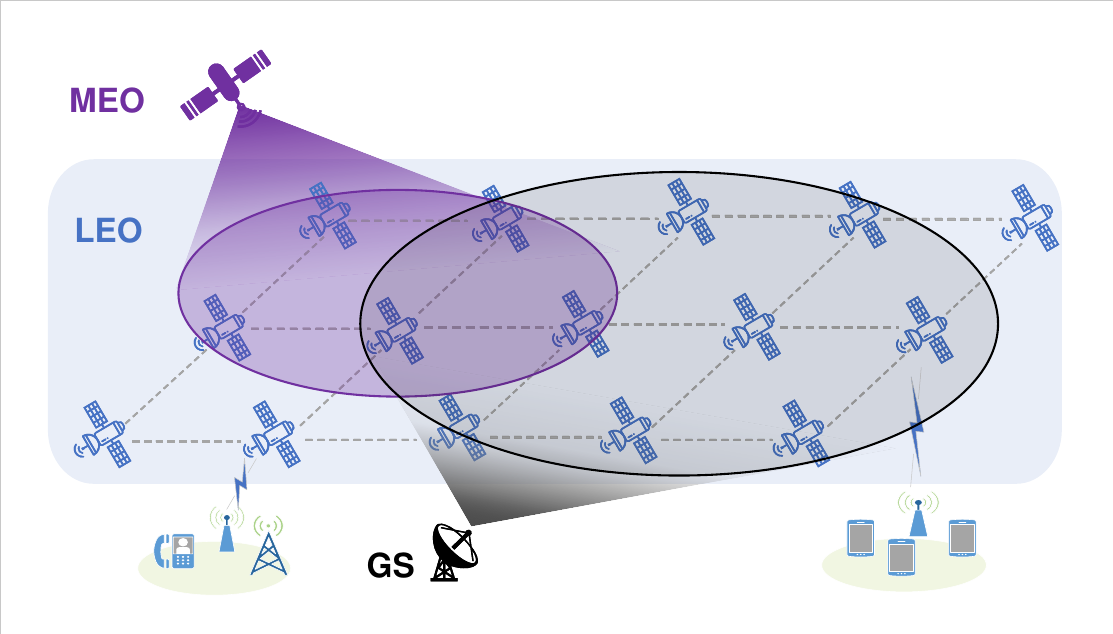}
  \caption{Unified hybrid control plane architecture combining ground stations and MEO satellites to manage LEO satellite domains with field-of-view constraints.}
    \label{system_scenario}
\end{figure}

To address these limitations, this paper introduces a comprehensive control overhead model that captures packet-level transmission costs and proposes Eunomia, a three-step domain partitioning framework that leverages predictable satellite directional mobility patterns within FOV constraints to achieve stable, movement-aware control domains. As illustrated in \figref{system_scenario}, our approach employs a unified hybrid control plane based on GS and MEO satellites that can achieve shorter control delays and complete LEO satellite coverage compared to traditional approaches.

\subsubsection{\textbf{FOV Constraints}} 
\begin{figure}[t]
  \centering
  \includegraphics[width=1.8in]{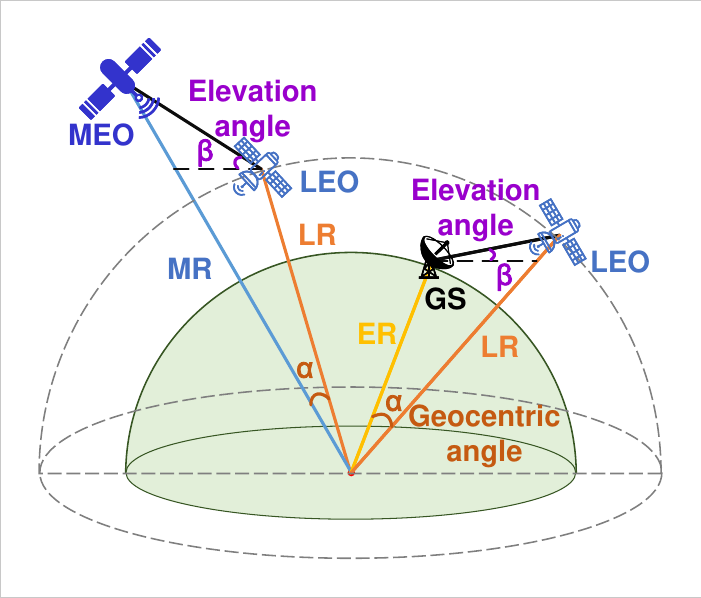}
  \caption{Geometric relationship between MEO controller and LEO satellite showing the elevation angle $\beta$ derived from geocentric angle $\alpha$, illustrating field-of-view coverage constraints in satellite networks.}
  \label{sat_perspective}
\end{figure}

The coverage area of each controller is fundamentally constrained by its FOV, determined by antenna beamwidth and elevation angle limitations. For MEO-LEO links, practical systems typically require a minimum elevation angle of $\pm 40^\circ$~\cite{plastikovHighgainMultibeamBifocal2016}, while GS controllers can communicate with LEO satellites across wider elevation ranges ($\pm 90^{\circ}$)~\cite{al-hraishawiSurveyNongeostationarySatellite2023}. As shown in \figref{sat_perspective}, the geocentric angle $\alpha$ between a LEO and MEO satellite determines whether the LEO falls within the controller's FOV. When the elevation angle $\beta$ (derived from $\alpha$ using trigonometric transformations) exceeds the antenna threshold, the LEO satellite lies outside the controller's coverage area.

\subsubsection{\textbf{Time Slot Model}} Due to orbital mechanics, FOV coverage is inherently dynamic. To manage this temporal variation, we employ a time slot model where each time slot $T_t$ represents a period during which the network topology remains stable. This assumption is validated by operational systems: LEO satellites follow predictable trajectories enabling stable control periods of 30-60 seconds, as demonstrated by Starlink's 15-second routing update intervals~\cite{tanveerMakingSenseConstellations2023, panMeasuringSatelliteLinks2024}. Controller migration occurs when elevation angles exceed FOV thresholds, triggering transition to the next time slot.

\subsubsection{\textbf{Control Overhead Sources}} The uneven distribution of satellite traffic within each time slot necessitates frequent controller scheduling for load balancing, introducing significant control overhead that domain partitioning strategies minimize through efficient traffic aggregation and adaptive routing. The allocation of LEO satellites within hybrid control planes faces new challenges due to antenna coverage overlap and satellite dynamics, necessitating sophisticated domain partitioning strategies.

\section{Overview of Eunomia} \label{sec_ove}

\begin{figure*}[t]
  \centering
  \includegraphics[width=5in]{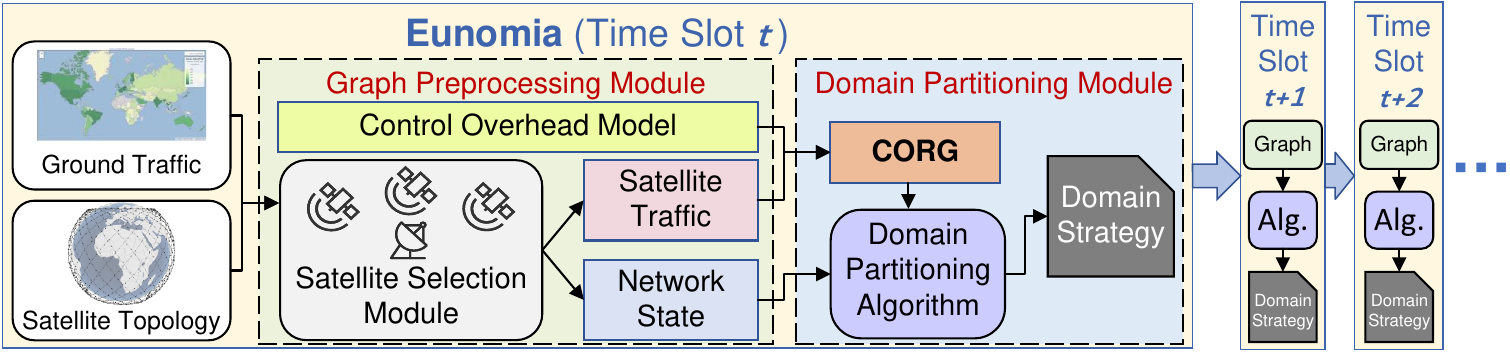}
  \caption{The overview workflow of Eunomia. Firstly, traffic preprocessing and CORG generation using satellite network information from current time slot. Then spectral clustering-based domain partitioning with controller assignment. Each subsequent time slot inherits and optimizes strategies from previous time slots to maintain domain stability while adapting to satellite movement.}
  \label{fig_overview}
\end{figure*}

To address the aforementioned challenges, we propose Eunomia, a domain partitioning framework. As illustrated in \figref{fig_overview}, Eunomia's workflow consists of three phases: the traffic preprocessing maps terrestrial demands onto the satellite network, CORG generation transforms network topology into control overhead relationships, and the partitioning algorithm generates the final domain strategy.

Eunomia's architecture consists of two main modules that work sequentially to achieve optimal domain partitioning:

\subsubsection{\textbf{Graph Preprocessing Module}} This module gathers satellite network state information and transforms it into actionable metrics for domain partitioning. During time slot $T_t$, while the network topology remains stable, the traffic distribution is unknown. We use the traffic distribution from the previous time slot as input, justified by traffic continuity and the temporal stability of ground patterns. The control overhead model (detailed in \secref{sec_mod}) quantitatively analyzes control message transmission costs, ultimately producing CORG that reflects control cost relationships between satellites.

\subsubsection{\textbf{Domain Partitioning Module}} This module utilizes network graphs and CORG to perform domain partitioning through Eunomia's three-step algorithm: FOV domain determination; overlapping region clustering using spectral clustering; and controller-cluster matching and movement-aware boundary fine-tuning. The algorithm can be executed in a distributed manner, requiring only coordination with adjacent controllers, which enables isolation of small-scale domain updates and enhances the strategy's effectiveness.

In the domain strategy model, binary variables \( x_{i,j}^t \) and \( y_{d,k}^t \) represent control relationships: \( x_{i,j}^t = 1 \) if satellites \( i \) and \( j \) belong to the same domain during \( T_t \), and \( y_{d,k}^t = 1 \) if domain \( d \) is managed by controller \( k \).

\section{Control Overhead Model}\label{sec_mod}
This section models the network framework and outlines the optimization objectives for domain partitioning.

\subsection{Hierarchical Satellite Network Model}

During a topology-stable time slot, the hierarchical satellite network graph comprises MEO satellites, LEO satellites, and ground stations as nodes, with inter-satellite and satellite-ground links as edges. Controllers (GS and MEO satellites) possess computational capabilities for route computation and network management, with GS controllers typically more powerful than MEO controllers. LEO satellites function as transparent data forwarding switches, maintaining flow tables that record processing rules for different traffic flows.

\subsection{Control Message Transmission Overhead Components}

The total control overhead $W_{\text{CTL}}(t)$ that impacts domain partitioning consists of three main components:

\begin{equation}
\label{equ_total_control_overhead}
W_{\text{CTL}}(t) = W_{\text{FLOW}}(t) + W_{\text{SYNC}}(t) + W_{\text{MIG}}(t)
\end{equation}
that represent the cost of flow table updates, domain synchronization, and controller migration, respectively.
All overheads are expressed in time to ensure unit consistency.

\subsubsection{\textbf{Flow Table Update Overhead}}
When a LEO switch receives an unknown flow, it must request flow table updates from its domain controller. This overhead can be classified into intra-domain requests (source and destination within the same domain) and inter-domain requests (spanning multiple domains). The flow table update overhead is:

\begin{equation}
\label{equ_flow_overhead}
W_{\text{FLOW}}(t) = \sum_{i,j \in V} \lambda_{i,j}^t \cdot \left( \frac{M_{\text{fl}}}{B_{i,j}} + D_{\text{prop}}(i,j) \right) \cdot h_{i,j}
\end{equation}
where $\lambda_{i,j}^t$ is the flow arrival rate between satellites $i$ and $j$; $M_{\text{fl}} = 36$ bytes is the flow table update message size; $B_{i,j}$ is available bandwidth; $D_{\text{prop}}(i,j)$ is propagation delay based on inter-satellite distance; and $h_{i,j}$ represents the number of hops required for the control message to travel from satellite $i$.

\subsubsection{\textbf{Domain Synchronization Overhead}}
Domain synchronization involves intra-domain synchronization (LEO switches reporting link states to controllers) and inter-domain synchronization (controllers coordinating global network views). The synchronization overhead is:

\begin{equation}
W_{\text{SYNC}}(t) = W_{\text{SYNC}}^{\text{in}}(t) + W_{\text{SYNC}}^{\text{out}}(t)
\end{equation}
where intra-domain synchronization overhead is:
\begin{equation}
\label{equ_sync_intra}
W_{\text{SYNC}}^{\text{in}}(t) = \sum_{d \in D} f_{\text{sync}} \cdot \max_{i \in d} \left( \frac{|E_d| \cdot M_{\text{sync}}}{B_{i,k_d}} + D_{\text{prop}}(i,k_d) \right)
\end{equation}

and inter-domain synchronization overhead is:
\begin{equation}
\label{equ_sync_inter}
W_{\text{SYNC}}^{\text{out}}(t) = f_{\text{sync}} \cdot \max_{k \in K} \sum_{\substack{k' \in K, \\ k' \neq k}} \! \left( \frac{|D_k| \cdot M_{\text{sync}}}{B_{k,k'}} + D_{\text{prop}}(k,k') \right)
\end{equation}

 $f_{\text{sync}}$ is synchronization frequency, that is, the number of times it is synchronized within a unit of time (one second); $|E_d|$ represents edges within domain $d$; $M_{\text{sync}} = 24$ bytes is the synchronization message size per edge; and $|D_k|$ represents the size of domain managed by controller $k$.

\subsubsection{\textbf{Controller Migration Overhead}}
Due to satellite mobility and FOV changes, controllers must periodically migrate domain control responsibilities. The migration overhead accounts for both state transfer and handover notification costs:

\begin{equation}
\label{equ_migration_overhead}
W_{\text{MIG}}(t) = \sum_{d \in D} f_{\text{mig}}^d \cdot \left( W_{\text{st}}^d + W_{\text{ho}}^d \right)
\end{equation}
where $f_{\text{mig}}^d$ is the migration frequency for domain $d$. State transfer cost $W_{\mathrm{st}}^d$ is the total volume of controller state to be synchronized during migration (e.g., flow tables and domain metadata) divided by the available transmission bandwidth. Handover notification cost $W_{\mathrm{ho}}^d$ is computed from the number of satellites requiring the control handover message, the size of each notification, and the per-satellite processing effort. 

\subsection{Path Computing Overhead}

Path computation overhead $W_{\text{CPT}}(t) = W_{\text{CPT}}^{\prime}(t) + W_{\text{CPT}}^{\prime\prime}(t)$ includes intra-domain and inter-domain computation:

\begin{equation}
\label{equ_cpt_delay}
\left\{
\begin{aligned} 
& W_{\text{CPT}}^{\prime}(t) = \sum_{d \in D} \sum_{k \in K} \frac{\mathsf{f}(|V_d|)}{C_k} \cdot F_{\text{prop}}^{d,t} \cdot y_{d,k}^t \\
& W_{\text{CPT}}^{\prime\prime}(t) = \sum_{d \in D} \sum_{k \in K} \frac{\mathsf{f}(|D|)}{C_k} \cdot F_{\text{inter}}^{d,t} \cdot y_{d,k}^t \\
\end{aligned}
\right.
\end{equation}
where $C_k$ is controller capacity, $F_{\text{prop}}^{d,t}$ and $F_{\text{inter}}^{d,t}$ are request counts, and $\mathsf{f}(\cdot)$ is a convex function related to algorithm complexity.

\subsection{Problem Formulation} \label{problem_formulation}
The Domain Partitioning Problem (DPP) minimizes total control and computation overhead:

\begin{equation}  
\label{equ_problem} 
\mathbf{P: }  \min \sum_{t \in T} \left( W_{\text{CTL}}(t) + \lambda W_{\text{CPT}}(t) \right)
\end{equation}
subject to constraints ensuring each domain has one controller in~\equref{cons:domain_control}, each satellite belongs to one domain in~\equref{cons:satellite_assignment}, connectivity within domains in~\equref{cons:connectivity}, FOV limitations in~\equref{cons:fov}, and binary assignment variables in~\equref{cons:binary}. The parameter $\lambda$ balances control overhead and path computation costs in~\equref{equ_problem}.

\begin{align}
\sum_{k \in K} y_{d,k}^t &= 1, \quad \forall d \in D, \forall t \in T \label{cons:domain_control} \\
\sum_{j \in V} x_{i,j}^t &= |D_i^t| - 1, \quad \forall i \in V, \forall t \in T \label{cons:satellite_assignment} \\
x_{i,j}^t &\leq \text{Connected}(i,j), \quad \forall i,j \in V, \forall t \in T \label{cons:connectivity} \\
y_{d,k}^t &\leq \text{FOV}(k,d), \quad \forall d \in D, \forall k \in K, \forall t \in T \label{cons:fov} \\
x_{i,j}^t, y_{d,k}^t &\in \{0,1\}, \quad \forall i,j \in V, \forall d \in D, \forall k \in K, \forall t \in T \label{cons:binary}
\end{align}

The DPP is NP-hard due to non-convex terms and discrete FOV constraints, justifying our heuristic three-step approach.

\subsection{Performance Metrics}
To evaluate domain partitioning effectiveness, we define key performance metrics that capture system behavior under different operational conditions:

The fraction of flow establishment requests that cannot be processed due to controller overload or network congestion:
\begin{equation}
\label{equ_drop_rate}
R_{\text{drop}}(t) = \frac{\text{Number of dropped requests during } T_t}{\text{Total number of requests during } T_t}
\end{equation}

This metric reflects the system's ability to handle peak traffic loads and indicates when controllers approach their processing capacity limits.

The ratio of useful control operations to total control overhead quantifies how effectively the control plane utilizes network resources:
\begin{equation}
\label{equ_efficiency}
\eta_{\text{control}}(t) = \! \frac{W_{\text{FLOW}}(t)}{W_{\text{CTL}}(t)} = \! \frac{W_{\text{FLOW}}(t)}{W_{\text{FLOW}}(t) + W_{\text{SYNC}}(t) + W_{\text{MIG}}(t)}
\end{equation}

Higher efficiency indicates that more control overhead is devoted to serving user requests, rather than maintaining network state, which is crucial for evaluating the quality of domain partitioning strategies.

\section{Domain Partitioning Algorithm}\label{sec_alg}

\begin{figure*}[t]
  \centering
  \includegraphics[width=6.2in]{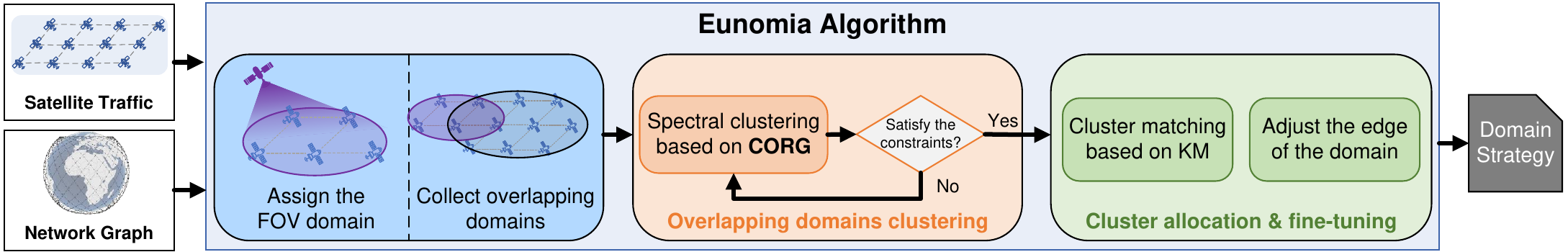}
  \caption{Flowchart of the domain partitioning algorithm. Firstly, FOV domain determination. Secondly, spectral clustering for overlapping regions. At last, controller-cluster matching with mobility-aware boundary fine-tuning.}
  \label{domain_alg}
\end{figure*}

To address the NP-hard domain partitioning problem, Eunomia employs a three-step approach that progressively narrows the solution space: determine FOV domains for each controller based on elevation angle constraints; cluster overlapping regions using spectral clustering to minimize control overhead; and match clusters to controllers with mobility-aware boundary fine-tuning. The algorithm workflow is shown in \figref{domain_alg}.

\begin{figure}[t]
  \centering
  \includegraphics[width=3.1in]{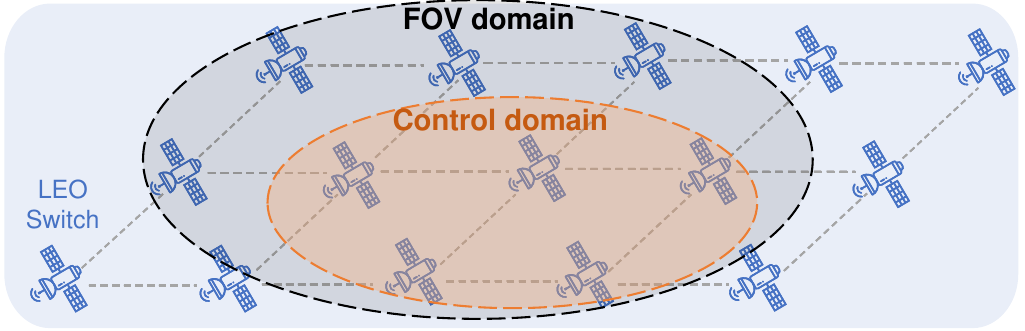}
  \caption{Containment relationship between FOV domains (determined by antenna elevation angle constraints) and control domains (actual satellites managed by controllers) in hierarchical satellite networks.}
  \label{visual_ctr_domain}
\end{figure}

\subsection{FOV and Control Domain Determination}

In hierarchical satellite network architecture, we distinguish between two types of domains: the FOV domain encompasses all LEO satellite switches within a controller's line of sight (determined by elevation angle constraints), while the control domain represents the actual satellites managed by that controller (\figref{visual_ctr_domain}). Due to physical limitations, control domains must be subsets of FOV domains.

The FOV domain size is determined by the controller's beam coverage, which may lead to overlaps between adjacent controllers' FOV domains. Each controller's FOV domain can be divided into non-overlapping and overlapping regions. LEO satellites in non-overlapping regions can only communicate with a single controller, making domain assignment straightforward. In Step 1, we directly assign satellites in non-overlapping regions to their respective controllers. The challenge lies in allocating satellites in overlapping regions, which requires the subsequent two steps.

\subsection{Spectral Clustering for Overlapping Regions}

For overlapping FOV regions where multiple controllers compete for satellite control, we apply spectral clustering based on the Control Overhead Relationship Graph (CORG). The CORG transforms network topology into control cost relationships, where edge weights $\xi_{i,j}$ between satellites $i$ and $j$ capture the comprehensive control overhead:

\begin{equation}
\xi_{i,j}\! =\! \alpha \cdot  W_{\text{FLOW}}\!(i,j)\! +\! \beta \cdot W_{\text{SYNC}}(i,j)\! +\! (1\!-\!\alpha\! -\! \beta) W_{\text{MIG}}(i,j)
\end{equation}
where $W_{\text{FLOW}}(i,j)$ represents flow table update costs, $W_{\text{SYNC}}(i,j)$ denotes synchronization overhead, and $W_{\text{MIG}}(i,j)$ captures mobility-induced migration costs. The weighting parameters ($\alpha = 0.5$, $\beta = 0.3$) are empirically determined to prioritize flow control efficiency while considering synchronization and mobility factors.

\begin{algorithm}[t]
\caption{Spectral Clustering for Domain Partitioning}
\label{alg:SpectralClustering}
\renewcommand{\algorithmicrequire}{\textbf{Input:}}
\renewcommand{\algorithmicensure}{\textbf{Output:}}
\begin{algorithmic}[1]

\REQUIRE $G_{\text{ovp}}^t$ (overlapping subgraph), $m$ (number of clusters equal to controllers in region)
\ENSURE $m$ LEO clusters $\{C_i\}_{i=1}^m$

\STATE \textbf{//} \textit{Build similarity matrix using Gaussian kernel}
\STATE Construct similarity matrix:
\STATE \quad $\displaystyle \xi(x_i, x_j) \gets
      \exp\Bigl(-\frac{\xi_{i,j}}{2\sigma^2}\Bigr),$ $\forall\,x_i, x_j \in G_{\mathrm{ovp}}$

\STATE \textbf{//} \textit{Compute normalized graph Laplacian}
\STATE $L \gets \mathrm{Laplacian}(\Xi)$,\quad $\hat{L} \gets \mathrm{Normalize}(L)$

\STATE \textbf{//} \textit{Extract eigenvectors for spectral embedding}
\STATE $U_m \gets \mathrm{Eigenvectors}(\hat{L}, m)$ \textit{//m smallest eigenvalues}

\STATE \textbf{//} \textit{Normalize rows for k-means input}
\STATE $y_i \gets u_i / \|u_i\|,\quad \forall\,u_i \in U_m$

\STATE \textbf{//} \textit{Apply k-means clustering in spectral space}
\STATE $\{C_i\}_{i=1}^m \gets k\text{-means}(\{y_i\}_{i=1}^n)$

\IF{Multiple virtual controller nodes in same cluster}
    \STATE Remove highest-weight edges, rerun k-means \textit{//Ensure cluster validity}
\ENDIF

\end{algorithmic}
\end{algorithm}

Spectral clustering groups together satellites that share similar control overhead patterns, effectively partitioning the overlapping region into candidate control domains. The number of clusters $m$ equals the number of controllers competing for the overlapping region.

\subsection{KM-Based Controller Matching and Boundary Fine-tuning}

The final step employs the Kuhn-Munkres (KM) algorithm to optimally match clusters to controllers while ensuring connectivity and minimizing total assignment cost. The cost matrix is constructed using geographic distances between cluster centroids and controller positions, ensuring that nearby clusters are assigned to proximate controllers to maintain domain connectivity.

To reduce domain churn caused by satellite motion, we propose an edge fine-tuning mechanism that dynamically updates control-domain boundaries using predictable LEO trajectories. First, we group satellites by flight direction (e.g. northbound vs. southbound) and adjust boundary assignments to limit future inter-domain handovers. Then, by clustering satellites with similar velocity vectors relative to their controllers, our method yields control domains that remain stable over time and naturally follow the underlying orbital motion.

This fine-tuning process is particularly effective in hierarchical satellite networks, where LEO satellites often move in predictable patterns relative to their controllers. By aligning control domains with these movement patterns, we can reduce the frequency of domain migrations, leading to more stable and efficient network operation.
This fine-tuning process reduces the frequency of domain migrations by up to 40\% compared to static assignment approaches, as satellites moving in the same direction tend to remain within the same controller's FOV for longer periods.

\subsection{Algorithm Complexity Analysis}

To provide theoretical guarantees for Eunomia's computational efficiency, we analyze the time complexity of each algorithmic component and evaluate algorithm performance.

\subsubsection{\textbf{FOV Domain Determination}} Computing the FOV domain for each controller requires evaluating elevation angles between all controller-satellite pairs. Given $|K|$ controllers and $|V|$ LEO satellites, this step has complexity $O(|K| \cdot |V|)$. Since $|K| \ll |V|$ in practical satellite networks (typically $|K| \approx 10-50$ while $|V| \approx 1000-10000$), this step is computationally efficient.

\subsubsection{\textbf{Spectral Clustering}} The spectral clustering component dominates the computational complexity. For an overlapping region with $n_{\text{ovp}}$ satellites and $m$ controllers:
\begin{itemize}
\item Similarity matrix computation: $O(n_{\text{ovp}}^2)$.
\item Eigenvalue decomposition: $O(n_{\text{ovp}}^3)$ using standard algorithms.
\item K-means clustering: $O(m \cdot n_{\text{ovp}} \cdot I_{\text{kmeans}})$ where $I_{\text{kmeans}}$ is the number of iterations.
\end{itemize}

Since overlapping regions are typically small compared to the entire network ($n_{\text{ovp}} \ll |V|$), and eigenvalue decomposition can be accelerated using sparse matrix techniques when the graph connectivity is low, the practical complexity is manageable.

\subsubsection{\textbf{KM-based Matching}} The Kuhn-Munkres algorithm for optimal assignment has complexity $O(m^3)$ where $m$ is the number of clusters (controllers) in the overlapping region. Since $m$ is bounded by the number of controllers in each overlapping area (typically $m \leq 5$), this step is computationally inexpensive.

The total time complexity of Eunomia is:

\begin{equation}
\label{equ_complexity}
T_{\text{Eunomia}} = O\left(|K| \cdot |V| + \sum_{r \in R_{\text{ovp}}} (n_r^3 + m_r^3)\right)
\end{equation}
where $R_{\text{ovp}}$ represents all overlapping regions, $n_r$ is the number of satellites in region $r$, and $m_r$ is the number of controllers in region $r$.

Importantly, this complexity scales much better than exhaustive domain partitioning approaches, which would require $O(|K|^{|V|})$ complexity to evaluate all possible assignments. Since $\sum_{r \in R_{\text{ovp}}} n_r \ll |V|$ (overlapping regions cover only a fraction of total satellites), Eunomia achieves near-linear scaling in practice.

Compared to benchmark algorithms:
\begin{itemize}
\item AROA~\cite{liOptimizedControllerProvisioning2023}: $O(|V|^2 \log |V|)$ due to regularization and randomized rounding.
\item ILA~\cite{huangEfficientDifferentiatedRouting2024}: $O(|V|^3)$ for branch-and-bound optimization.
\item CCPA~\cite{chenMobilityLoadadaptiveController2021}: $O(|V|^2)$ for greedy controller placement.
\end{itemize}

Eunomia's complexity is substantially lower, explaining the 77.7\% runtime reduction observed in our experiments (\secref{sec_eva}).

\section{System Implementation}\label{sec_sys_imp}
To validate the performance of Eunomia, we implemented the system on the satellite Internet digital twin platform, Plotinus~\cite{gaoPlotinusSatelliteInternet2024}. This section provides an explanation of the workflow and technical details of the system implementation.

\subsection{Overview of Emulation System}
\begin{figure}[t]
  \centering
    \includegraphics[width=3.2in]{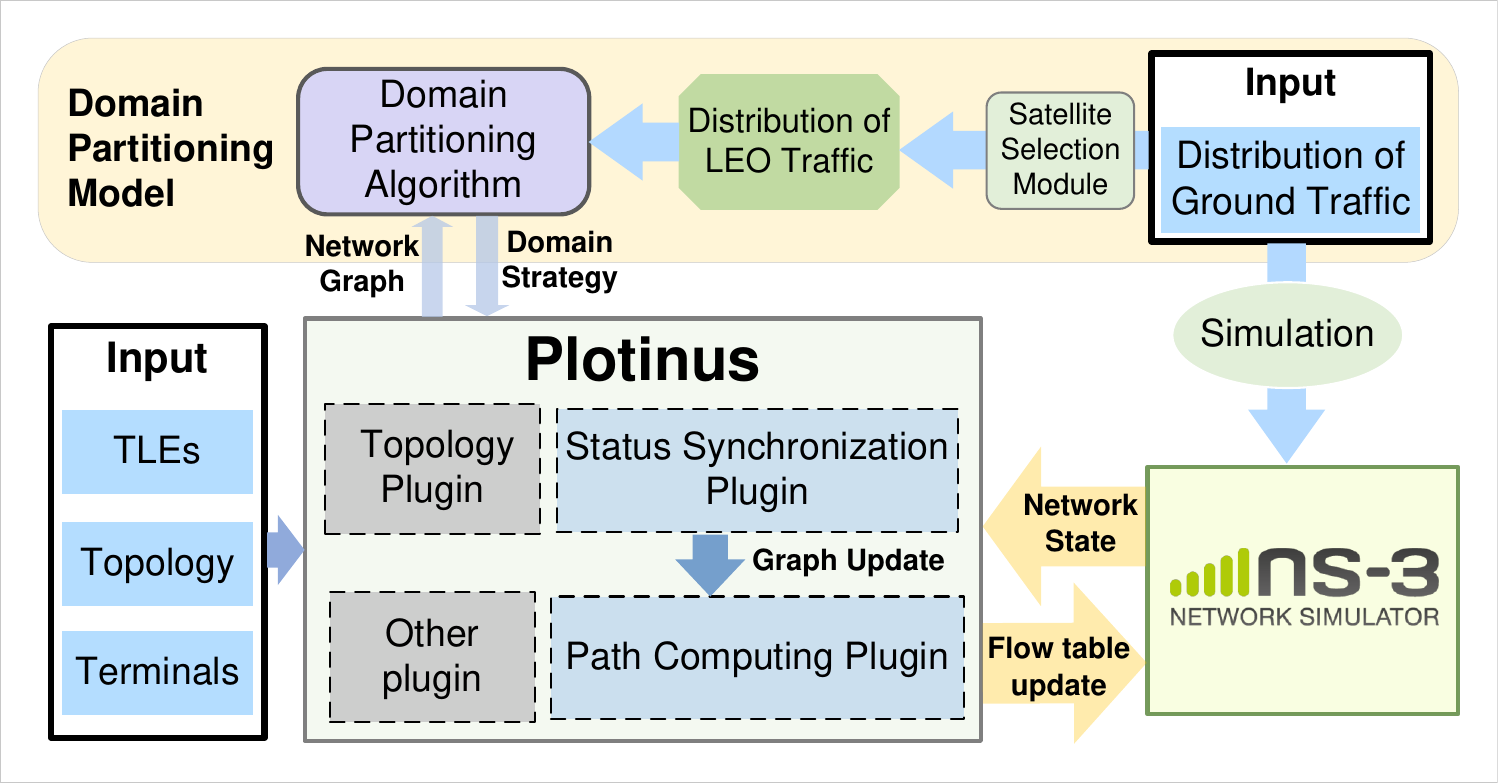}
    \caption{Extended Plotinus emulation platform architecture for hierarchical satellite network evaluation, integrating constellation modeling, traffic distribution, NS-3 simulation, and domain partitioning strategy assessment.}
  \label{plt_system}
\end{figure}

We extended Plotinus to build a comprehensive emulation system.
Plotinus is a digital twin system that supports end-to-end emulation of satellite Internet networks. We customize the architecture of the control plane and the communication model by adding microservices modules. The experimental framework of this study is illustrated in \figref{plt_system}, which includes two input modules: one for importing the constellation model and another for user traffic distribution. The constellation model is used to establish a virtual network model, while the user traffic distribution serves two purposes. First, it is imported into the NS-3 module to complete traffic emulation. Second, it is processed through a satellite selection module to convert it into LEO onboard traffic, which is then imported into the domain partitioning strategy model as a factor for evaluating the partitioning algorithm.

To enable Plotinus to support a distributed network management model, we activated OpenFlow switches in NS-3 and used the distributed Bellman-Ford algorithm for route computation. The controller communication is implemented using the Python-based ZeroMQ library~\cite{hintjensZeroMQMessagingMany2013}, which shares the global network topology via topology updates and computes intra-domain and inter-domain routes. 

\subsection{LEO Traffic Distribution}
\label{traffic_dstr}
Measuring traffic between LEO satellites within the hierarchical satellite network architecture presents significant challenges. Due to the proprietary nature of systems like Starlink, obtaining direct traffic distribution data from existing satellite networks is not feasible. Consequently, most existing literature relies on ground-based traffic mapping methods to estimate satellite traffic. For instance, in~\cite{huangEfficientDifferentiatedRouting2024}, the Earth's surface is divided into 72 regions, with Internet user data from 2021 scaled by \(10^6\) and mapped to these regions to represent satellite traffic. 
Similarly,~\cite{liOptimizedControllerProvisioning2023} using a gravity model to emulate satellite traffic loads.
However, these approaches feature a coarse granularity, suitable only for simulating small-scale satellite networks and inadequate for accurately reflecting LEO traffic dynamics.

To achieve more fine-grained emulation of LEO traffic distribution, we optimized the traffic mapping methodology. As derived from~\cite{markovitzLEOSatelliteBeam2022}, a single LEO satellite in the Starlink constellation at an altitude of 550~km has a beam coverage area of approximately $1.033 \times 10^6$~km$^2$. Given that the Earth's surface area is around $5.1 \times 10^8$~km$^2$, a minimum of 494 ground cells is required to ensure independent coverage by a single satellite. Thus, to balance granularity and computational tractability, the original ground cells in~\cite{huangEfficientDifferentiatedRouting2024} were subdivided into a $3 \times 3$ grid, resulting in 648 cells.

Subsequently, the problem of LEO traffic distribution is transformed into a ground cell traffic distribution problem. We utilize a gravity model based on population density~\cite{liOptimizedControllerProvisioning2023} to calculate the traffic distribution for each ground cell, including the source and destination cells. Next, we add a time axis into the traffic data, aligning it with geographical daylight patterns. Finally, the traffic distribution of ground cells is mapped onto the LEO network, resulting in a three-dimensional traffic matrix composed of time slots, source LEO switches, and destination LEO switches. This baseline traffic matrix is denoted by $M_\text{base}$, where each element $m_{t,s,d} \in M_\text{base}$ represents the traffic volume from a source satellite $s$ to a destination satellite $d$ during a specific time slot $t$.

\subsection{Data Design for Control Message}
In \secref{sec_mod}, the control messages in hierarchical satellite network are categorized into three main types: flow table request messages, flow table update messages, and edge synchronization messages.

\begin{figure}[t]
  \centering
    $\begin{array}{ccc}
        \hspace{-2mm}\includegraphics[width=3.2in]{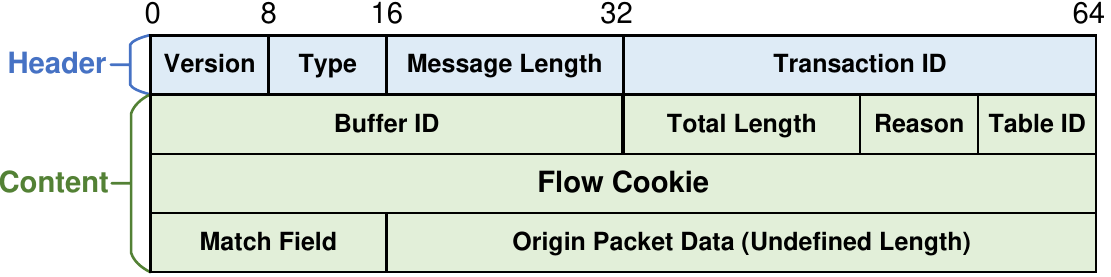}\\
        \hspace{0mm}\mbox{\footnotesize (a) \text{Flow table request message}}\\

        \hspace{-2mm}\includegraphics[width=3.2 in]{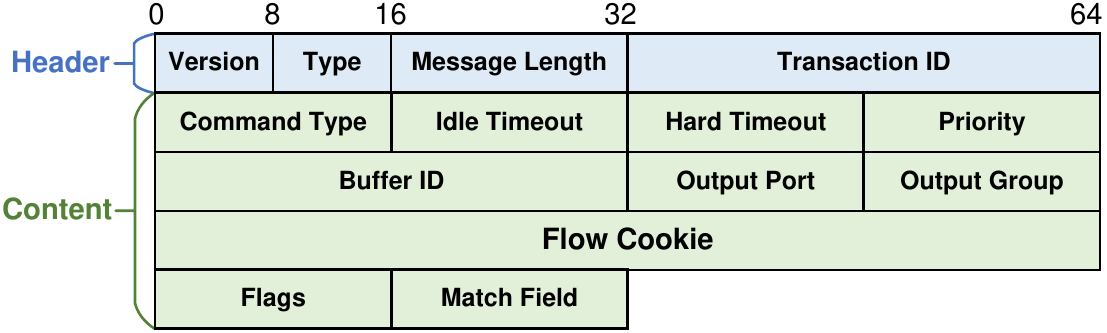}\\
        \hspace{0mm}\mbox{\footnotesize (b) \text{Flow table update message}}\\

        \hspace{-2mm}\includegraphics[width=3.2 in]{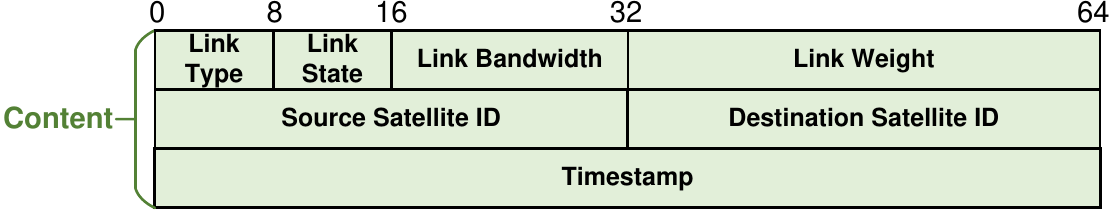}\\
        \hspace{0mm}\mbox{\footnotesize (c) \text{Edge synchronization message}}\\
    \end{array}$
    \caption{Control message formats in hierarchical satellite networks: (a) Flow table request message for unknown packet notifications, (b) Flow table update message (36 bytes) for routing instructions, and (c) Edge synchronization message (24 bytes) for topology updates.}
    \label{ctr_msg_abc}
\end{figure}

Flow table request messages are sent from LEO switches to the controller to notify the arrival of an unknown data packet and to request a flow table update. \figref{ctr_msg_abc}(a) shows the message format, the Header includes fields such as the version, which identifies the OpenFlow protocol version; type and length, which describe the message type and its length; and a Transaction ID for matching requests with responses. The Content section contains details such as the buffer address, total packet length, transaction reason, flow table ID, Cookie value, matching fields (including network type, IP address, and port number), and the initial packet's content. 

Flow table update messages are the responses from the controller after a flow table update request. The controller performs routing computations and then sends flow table update messages to the nodes along the optimal path. If the destination address is within the same domain, the controller directly updates the flow tables of these devices. For destinations in other domains, the controller collaborates with adjacent controllers to compute a global path and update the affected nodes. The format of a flow table update message is shown in~\figref{ctr_msg_abc}(b), the Content includes command type, idle timeout, maximum timeout, priority, buffer ID, output port, output group, flow Cookie, flags, and matching fields. A single flow table update message also totals $M_{\text{fl}} = 36$ bytes.

Edge synchronization messages are used to synchronize the status of an edge in the network graph and are crucial for topology updates. As illustrated in~\figref{ctr_msg_abc}(c), the format of an edge synchronization message includes link type, link status, bandwidth, weight, source and destination satellite IDs, and a timestamp of the update. The data required for a single message transmission amounts to $M_{\text{sync}} = 24$ bytes.

\section{Evaluation}\label{sec_eva}
In this section, we first introduce the emulation setup of the algorithm on the Plotinus system, followed by an analysis of the algorithm's performance under varying traffic scales, LEO constellation scales, and MEO controller scales.

\subsection{Emulation Setup}
We deployed Eunomia and benchmarks on the Plotinus platform for emulation. The entire Plotinus system is deployed on an Ubuntu server equipped with an Intel Xeon Gold 6418H \@2.10 GHz CPU and 256 GB of memory. 
The traffic model is based on the traffic matrix $M_\text{base}$ from \subsecref{traffic_dstr}, which is calibrated to represent a high-traffic scenario close to network saturation. 
To evaluate performance under varying degrees of load, we introduce a scaling factor $\gamma \in [0,1]$. The simulated traffic matrix is $M_\text{sim} = \gamma \cdot M_\text{base}$. This allows us to control the traffic intensity, ranging from no load ($\gamma = 0$) to a high-congestion state ($\gamma = 1$).

\begin{figure}[t]
  \centering
  \includegraphics[width=2in]{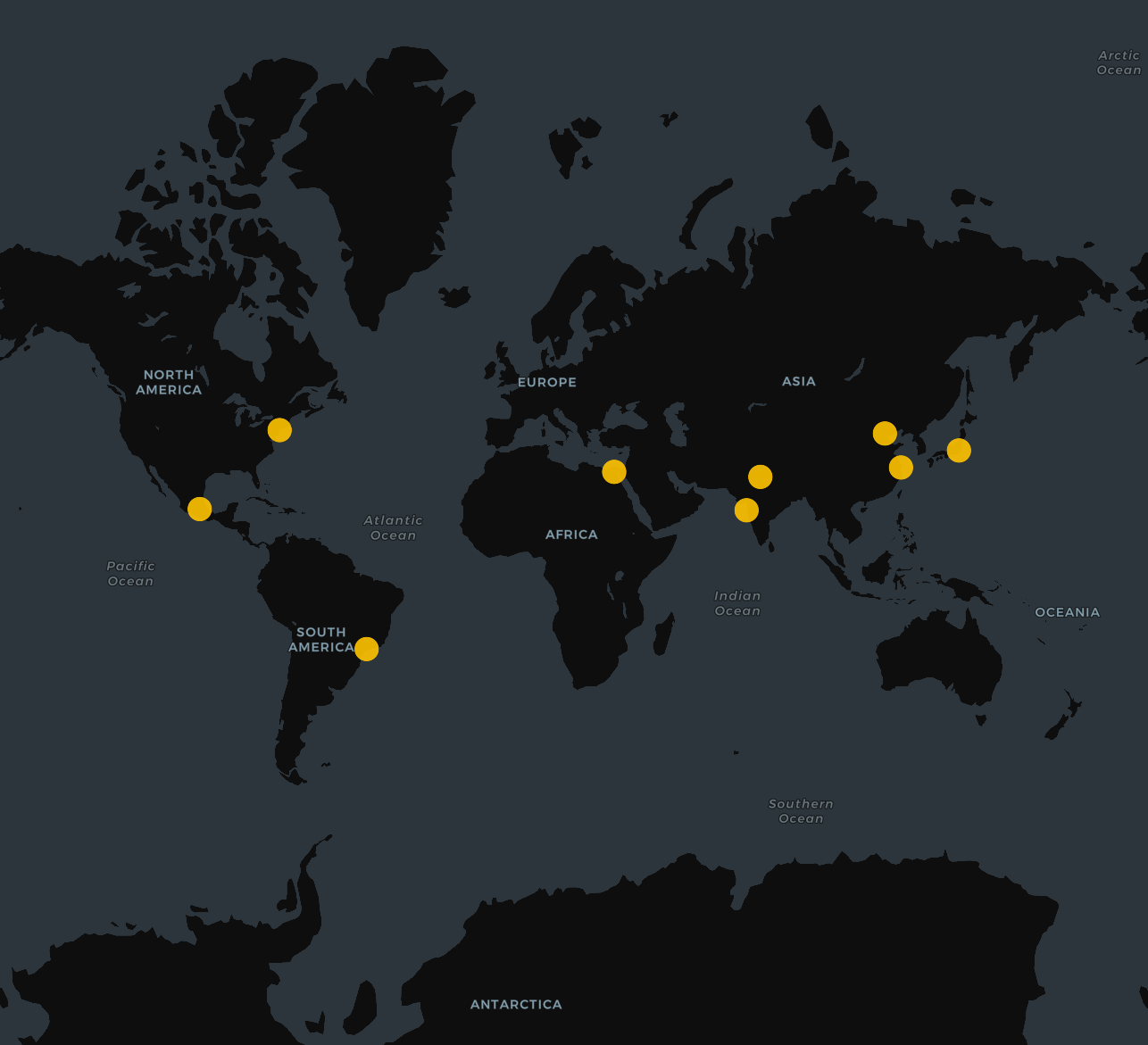}
  \caption{Distribution of ground stations in nine densely populated urban areas. The map shows the locations of ground stations in New York, London, Tokyo, Sydney, São Paulo, Cairo, Mumbai, Beijing, and Lagos.}
  \label{fig_gs_map}
\end{figure}

As shown in \figref{fig_gs_map}, we selected nine densely populated cities (New York, London, Tokyo, Sydney, São Paulo, Cairo, Mumbai, Beijing, Lagos) to represent the distribution of ground stations, chosen based on global Internet traffic distribution data~\cite{townsendNetworkCitiesGlobal2001}. This selection ensures geographic diversity and aligns with idealized satellite network deployment patterns.

FOV elevation thresholds are set to $\pm 40^\circ$ for MEO-LEO links and $\pm 90^\circ$ for GS-LEO links based on established satellite communication standards~\cite{plastikovHighgainMultibeamBifocal2016}. These conservative thresholds are validated by operational systems: Starlink terminals require minimum $25^\circ$ elevation~\cite{downsEffectsLowEarth2025}.

Processing capability ratios (GS : MEO : LEO = $10^5 : 10^2 : 1$) are based on hardware specifications from operational systems. Modern ground stations support processing capacities of 1-10 TFLOPS, MEO satellites achieve 10-100 GFLOPS~\cite{ortizOnboardProcessingSatellite2023}, while LEO satellites typically operate at 1-10 GFLOPS~\cite{huangIntegratedComputingNetworking2024}. Our capability ratios reflect these real-world constraints.

Unless otherwise stated, we used a default satellite network consisting of 8 MEO satellites at 10,354 km altitude and 1,584 LEO satellites at 550 km altitude, corresponding to Starlink's Shell 1 configuration. The MEO configuration uses 6 satellites in 2 orbital planes with $39.4^\circ$ inclination, providing global coverage while minimizing constellation complexity. Detailed parameters are provided in \tabref{tab_meo} and \tabref{tab_leo}.

We adopt the four most related and state-of-the-art algorithms as benchmarks.

\begin{itemize}
    \item \textit{CCPA}~\cite{chenMobilityLoadadaptiveController2021}: The CRG-based controller placement and assignment (CCPA) is designed specifically for deploying SDN controllers within LEO satellites, optimizing the placement dynamically. CCPA constructs a Control Graph Relationship (CGR) to analyze the load distribution among LEO satellites, employing a greedy candidate set to narrow the solution space.
    \item \textit{ILA} ~\cite{huangEfficientDifferentiatedRouting2024}:  The Inter-Layer Algorithm (ILA) is a domain partitioning algorithm managed by MEO satellites. ILA leverages a branch-and-bound approach to optimize domain sizes across multiple objectives, including maximum visibility time, transmission distance, and available connections for MEO-LEO inter-layer links.
    \item \textit{AROA}~\cite{liOptimizedControllerProvisioning2023}: The Approximate Regularization-based Online Algorithm (AROA) focuses on the hierarchical management of controllers across LEO satellites and GS. AROA calculates migration weights using regularized control overheads and obtains approximate solutions through online randomized rounding.
    \item \textit{ODC}~\cite{tanveerMakingSenseConstellations2023}: The One Domain Centralized (ODC) model is an industry-standard centralized single-domain management scheme inspired by Starlink's current management framework. ODC gathers satellite status data through distributed GS and aggregates it at a central GS for global network management.
\end{itemize}

\begin{table*}
  \centering
  \caption{MEO Constellation Models in Emulation}
  \label{tab_meo}
  \begin{tabular}{cccccc}
    \toprule
    Altitude (km)&Orbit Inclination&Eccentricity & Period (Min) &Total Number of Satellites & Number of Orbital Planes \\
    \midrule
    3000 & 63.4 & 0.1  & 150.46 & 36 &6\\
    6000 & 55 & 0.01 & 228.23 & 32 &4\\
    8070 & 53.1 & 0.001 & 287.93 & 20 &5 \\
    10354 & 39.4 & 0.0001 & 358.76 & 6 &2\\
  \bottomrule
\end{tabular}
\end{table*}

\begin{table*}
  \centering
  \caption{LEO Constellation Models in Emulation}
  \label{tab_leo}
  \begin{tabular}{cccccc}
    \toprule
    Constellation Name & Altitude (km)&Orbit Inclination &Total Number of Satellites & Number of Orbital Planes \\
    \midrule
    Iridium 780 ~\cite{fossaOverviewIRIDIUMLow1998} & 780 & 86.4 & 66 & 6   \\
    Telesat 1015 ~\cite{delportilloTechnicalComparisonThree2019} & 1015 & 98.98 & 351 & 27   \\
    OneWeb 1200 ~\cite{pachlerUpdatedComparisonFour2021}& 1200 & 87.9 & 720 & 18   \\
    Starlink 550 ~\cite{mcdowellLowEarthOrbit2020}& 550 & 53 & 1584 & 72  \\
    CSCN 365 ~\cite{xuSelforganizingControlMega2022}& 365 & 40 & 1848 & 33   \\
  \bottomrule
\end{tabular}
\end{table*}

\subsection{Performance Analysis Under Different Traffic Loads}
After completing a full satellite motion cycle, we assessed the performance of our proposed scheme against the benchmark algorithms under high-traffic load scenarios. In this network model, the MEO satellites fully cover all LEO satellites, enabling inter-layer link communication. 

\begin{figure}[t]
  \centering
    $\begin{array}{ccc}
        \hspace{0mm}\includegraphics[width=2.5 in]{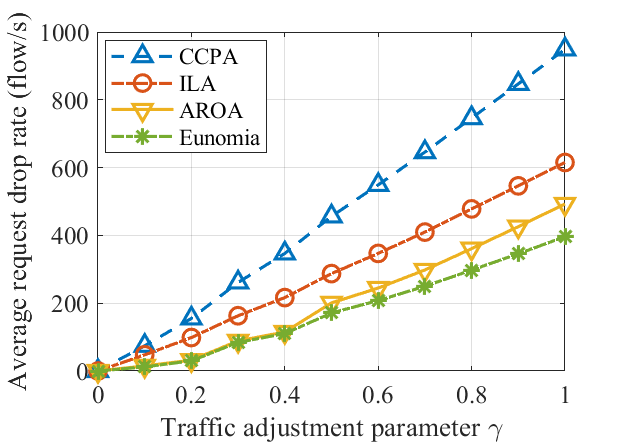} \\
        \hspace{0mm}\mbox{\footnotesize (a) \text{ Average request drop rate}}\\
        \hspace{0mm}\includegraphics[width=2.5 in]{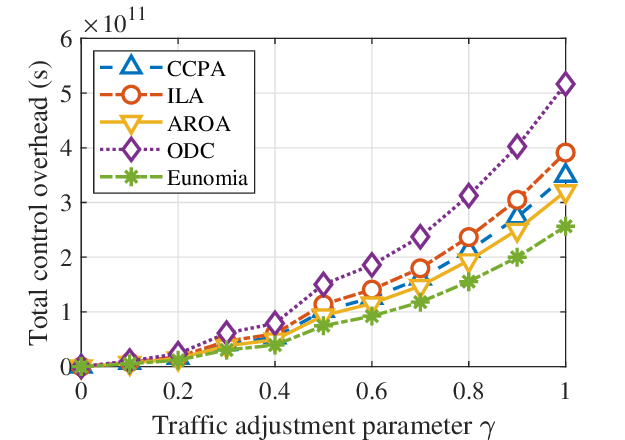}\\
        \hspace{0mm}\mbox{\footnotesize (b) \text{ Total control overhead time}}\\
    \end{array}$
    \caption{Performance comparison under varying traffic loads ($\gamma$).}
  \label{flow_1_2}
  % \vspace{-0.6cm}
\end{figure}

We first examined the impact of varying traffic loads on the request drop rate, as illustrated in \figref{flow_1_2}(a). Request drops in hierarchical satellite networks occur primarily for two reasons: either when controllers reach capacity limits and cannot process additional requests, or when network conditions make request resolution impossible during forwarding. Notably, even under low-load network scenarios, CCPA and ILA exhibit significantly poorer performance compared to the other three algorithms. The key difference lies in the use of GS as network controllers in the latter three algorithms. Consequently, in an ODC system where only GS serves as centralized controllers, it exhibits the lowest average request drop rate. In contrast, CCPA, constrained by the limited processing power of LEO satellites, shows an 89.6\% increase in the request drop rate under high-load conditions compared to ODC. With the MEO satellites enhancing the processing capabilities of GS, Eunomia achieves a 19.6\% lower average request drop rate under high-load scenarios compared to AROA and a 58.3\% reduction compared to CCPA.

Even under high-load scenarios, Eunomia maintains the lowest control overhead. As described in \secref{sec_mod}, control overhead is a critical metric for evaluating network control plane performance, reflecting the amount of control information required for resource management, request processing, and other operations. As shown in \figref{flow_1_2}(b), Eunomia reduces control overhead through CORG optimization process. Additionally, the direct connection between LEO switches and controllers in Eunomia saves substantial control message forwarding time, enabling Eunomia to outperform AROA. When the value of $\gamma$ is 1, Eunomia's total control overhead is 50.3\%, 34.4\%, 26.5\%, and 19.95\% lower than ODC, ILA, CCPA, and AROA, respectively.

\subsection{Performance Analysis Under Different LEO Constellations}
\begin{figure}[t]
  \centering
  \includegraphics[width=3.3in]{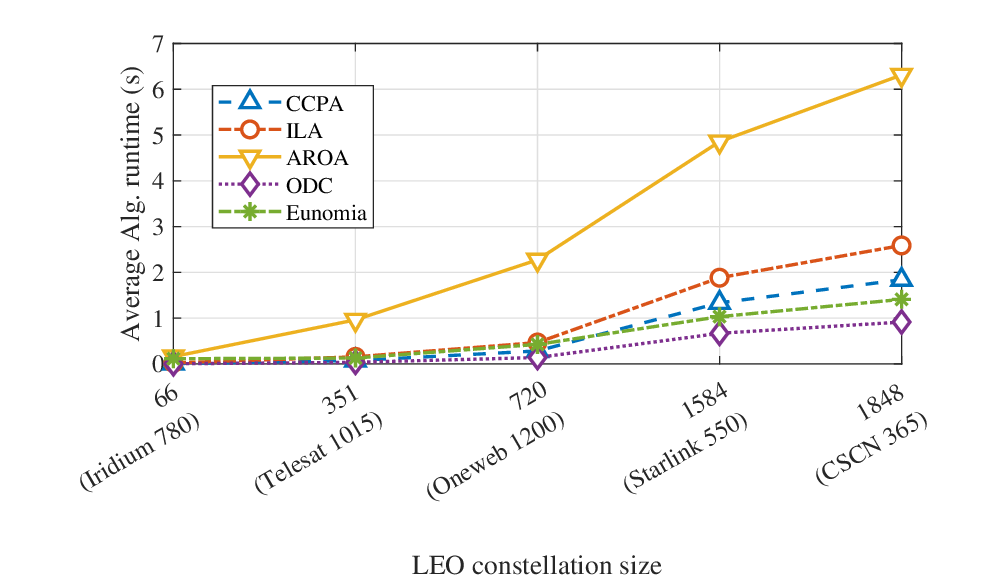}
  \caption{The algorithm runtime comparison across varying LEO constellation scales demonstrates that Eunomia achieves a 77.7\% average reduction in execution time compared to AROA.}
  \label{leo_3_1}
\end{figure}

\begin{figure*}[t]
  \centering
    $\begin{array}{ccc}
        \hspace{0mm}\includegraphics[width=2.29 in]{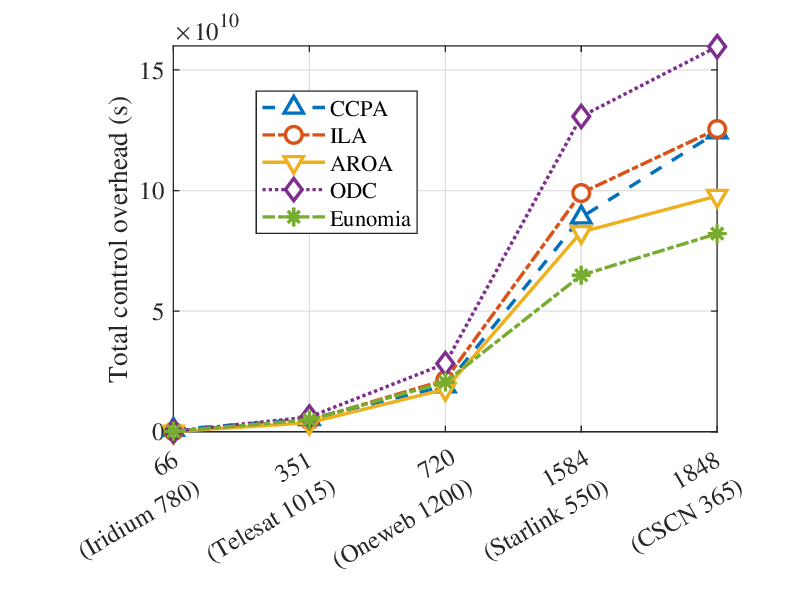}&
        \hspace{-1mm}\includegraphics[width=2.2 in]{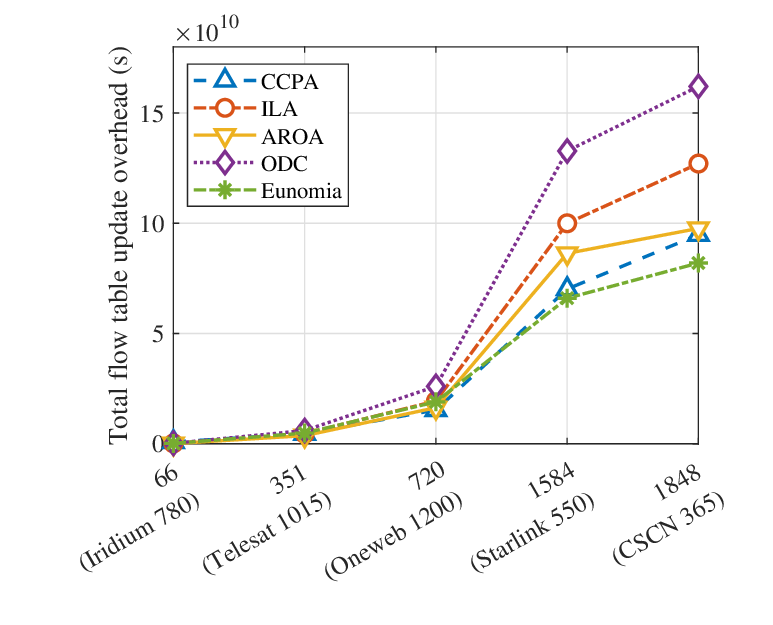}&
        \hspace{-2mm}\includegraphics[width=2.3 in]{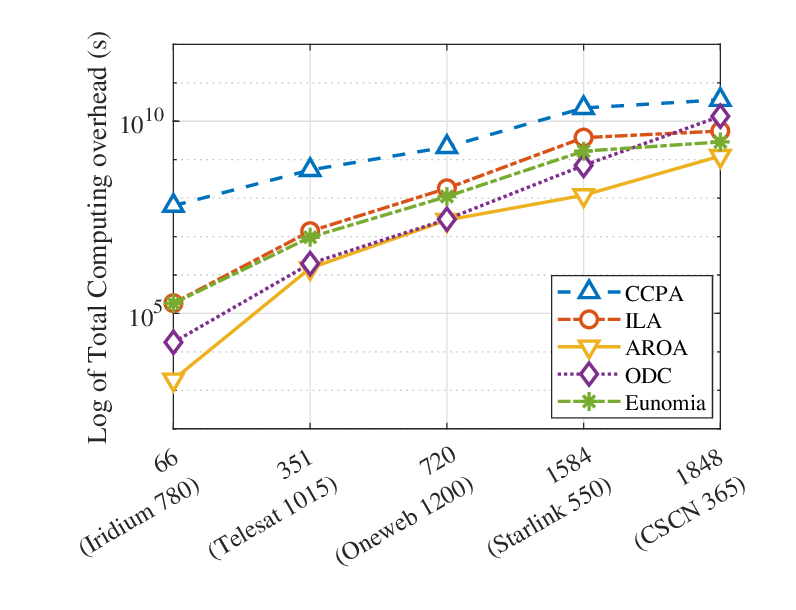} \\
        
        \hspace{0mm}\mbox{\footnotesize (a) \text{Total control overhead}}&
        \hspace{-1mm}\mbox{\footnotesize (b) \text{Total flow table update overhead}}&
        \hspace{-2mm}\mbox{\footnotesize (c) \text{Log of total computing overhead}}\\
    \end{array}$
    \caption{Performance evaluation across different LEO constellation sizes: (a) Total control overhead comparison, (b) Flow table update overhead analysis, and (c) Logarithmic scale computing overhead, demonstrating Eunomia's superior scalability in large-scale scenarios.}
  \label{fig_3_3}
\end{figure*}
We further evaluated the performance of the algorithms in multiple LEO constellation scenarios, with the parameters of these constellations provided in \tabref{tab_leo}. In the hierarchical satellite network framework, the time required to formulate a domain partitioning strategy is a critical performance metric, referred to as algorithm runtime. A shorter algorithm runtime indicates a better response to the dynamic nature of satellite networks, enhancing the effectiveness of the domain partitioning strategy. As shown in \figref{leo_3_1}, AROA improves the proportion of intra-domain traffic through a randomized rounding algorithm, but this process consumes a significant amount of time, leading to a decline in system performance. ODC, featuring a single large control domain, requires only one domain partitioning strategy under a stable topology, resulting in the shortest algorithm runtime. Benefiting from a top-down, three-step architecture and mobility optimization, Eunomia can efficiently maintain domain consistency and quickly adapt partitions between consecutive time slots. This makes Eunomia the second-fastest domain partitioning algorithm after ODC. In large-scale LEO constellation scenarios, Eunomia reduces the algorithm runtime by an average of 77.7\% compared to AROA and 45.5\% compared to ILA.

In large-scale LEO constellation scenarios, Eunomia exhibits lower total control overhead than other algorithms. As illustrated in \figref{fig_3_3}(a), the performance gap between different algorithms is minimal when the LEO constellation size is small. However, as the number of LEO satellites increases dramatically, the centralized architecture's control overhead escalates exponentially due to more frequent synchronization and slower request response times. As illustrated in \figref{fig_3_3}(b) and \figref{fig_3_3}(c), Eunomia,  demonstrates lower control overhead in large-scale LEO scenarios, due to its low flow table update overhead and routing computation time. Nevertheless, in smaller LEO constellations, such as those with 351 or 720 satellites, Eunomia's performance is slightly lower than AROA due to the higher number of control domains. This is because AROA and CCPA adaptively adjust the number of control domains, whereas Eunomia's domain count is tied to the number of MEO and GS nodes in the scenario. In larger LEO constellations, the bottleneck shifts from domain quantity to domain size, alleviating Eunomia's performance issues. This further validates Eunomia's effectiveness in large-scale scenarios of hierarchical satellite network.

\subsection{Scalability of Eunomia}

\begin{figure}[t]
  \centering
    $\begin{array}{ccc}
        \hspace{0mm}\includegraphics[width=2.8 in]{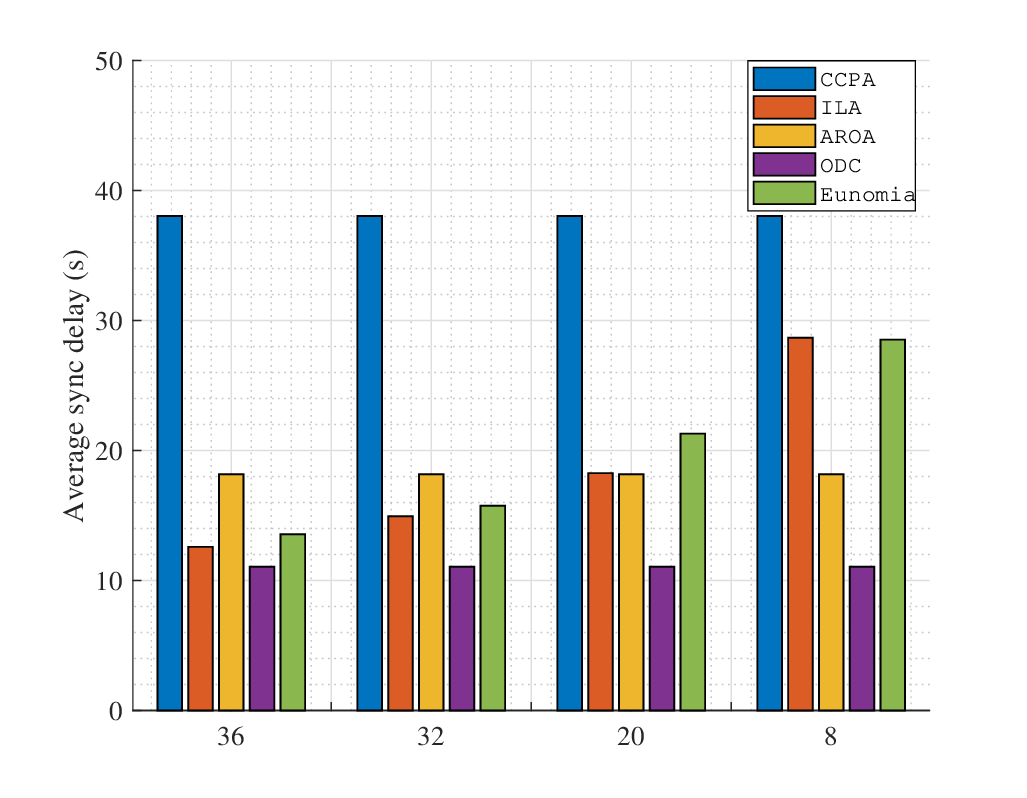}\\
         \hspace{0mm}\mbox{\footnotesize (a) \text{ Average \ synchronous  \ delay}}\\   
        \hspace{0mm}\includegraphics[width=2.8 in]{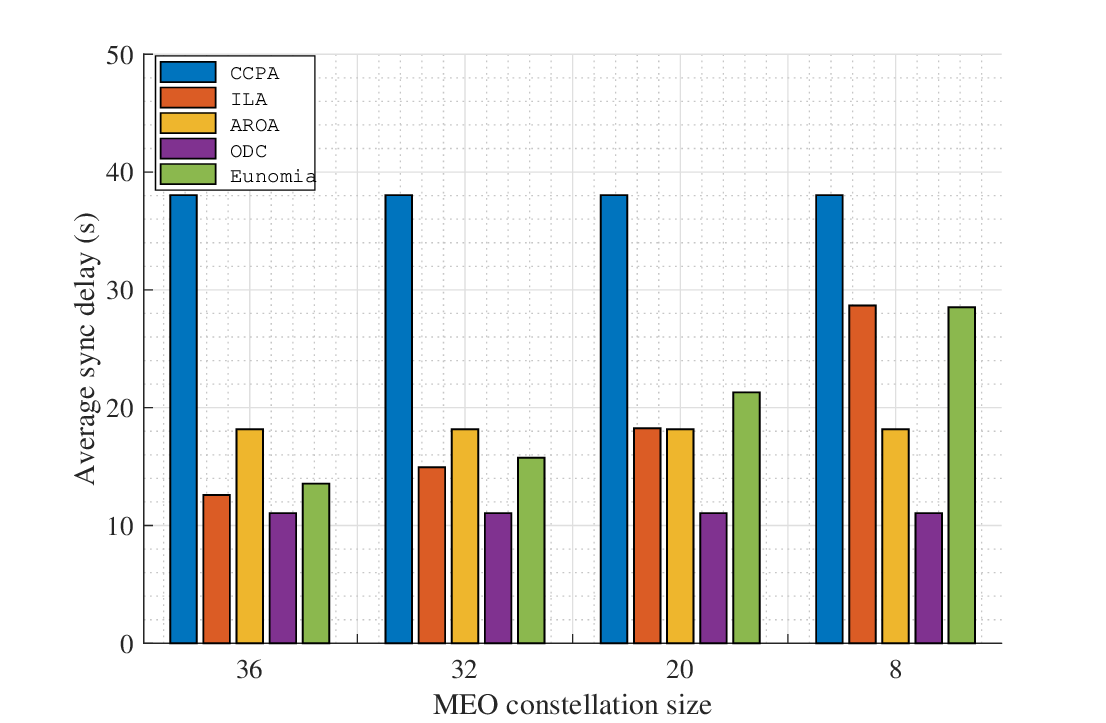}\\
        \hspace{0mm}\mbox{\footnotesize (b) \text{ Average \ response \ delay}}\\
    \end{array}$
    \caption{Scalability analysis under different MEO constellation configurations: (a) Average synchronization delay per domain showing the impact of MEO altitude on control plane performance, and (b) Average response delay per request demonstrating Eunomia's 51\% improvement over ILA.}
  \label{meo_2_2}
\end{figure}

To better demonstrate the scalability of Eunomia, we conducted algorithm evaluations under high traffic load scenarios using various MEO models listed in \tabref{tab_meo}. As shown in \figref{meo_2_2}, only ILA and Eunomia exhibit performance variation in relation to the MEO constellation size. \figref{meo_2_2}(a) illustrates the changes in average delay per domain synchronization, while \figref{meo_2_2}(b) shows the changes in average response delay per request. Both ILA and Eunomia experience increased average synchronization delays as the MEO constellation size grows; however, ILA's synchronization delay is slightly lower than Eunomia's in MEO constellations below an altitude of 10,000 km. This is because ILA maximizes the maintenance time of inter-layer links between MEO and LEO satellites, which aids in the state synchronization between MEO controllers and LEO switches. Nonetheless, ILA's failure to consider controller load balancing is one of its shortcomings, leading to a significantly higher average response delay compared to Eunomia. 

Average response time, defined as the average time required to complete a routing request, is one of the most critical factors influencing network performance. With the increase in MEO constellation altitude, Eunomia's average response time also increases due to the added propagation delay, but Eunomia still outperforms other algorithms. On one hand, Eunomia enhances traffic aggregation through CORG, effectively reducing the proportion of inter-domain traffic; on the other hand, Eunomia's intra-domain LEO satellites can communicate with the controller through ISL or satellite-to-ground links, eliminating the time required for inter-layer link forwarding. In \figref{meo_2_2}(b), the comparison between Eunomia and ILA underscores the importance of traffic optimization, while the comparison with AROA highlights the performance improvements brought by the MEO layer to hierarchical satellite network. Eunomia reduces the average response time by 51\% compared to ILA and by 17.6\% compared to AROA.

%-------------------------------------------------------------------------------
\section{Related Work}\label{sec_related}
%-------------------------------------------------------------------------------

The domain partitioning problem in satellite networks spans three major research areas: satellite network architectures, controller placement strategies, and mobility-aware network management. We review the evolution of these fields and position our contributions within the broader research landscape.

\subsection{Satellite Network Control Architectures}

Satellite network control has evolved through three distinct generations, each addressing scalability limitations of its predecessor. First-generation centralized approaches~\cite{maImplementationCentralizedControllers2022, tanveerMakingSenseConstellations2023} concentrate all control functions at ground stations, offering simplicity and global optimization at the cost of scalability and coverage limitations. While effective for small constellations (fewer than 100 satellites), these approaches struggle with the control overhead and latency requirements of mega-constellations.

Second-generation distributed approaches~\cite{liOptimizedControllerProvisioning2023, chenMobilityLoadadaptiveController2021} distribute control functions across LEO satellites themselves, improving scalability and reducing average control latency. However, these solutions impose significant computational and power constraints on LEO satellites and suffer from frequent control topology changes due to satellite mobility.

Third-generation hierarchical approaches~\cite{chenHierarchicalDomainBasedMulticontroller2022, xuDistributedMultilayerHierarchical2024} introduce intermediate control layers using GEO or MEO satellites. While addressing some scalability concerns, existing hierarchical solutions treat MEO satellites primarily as a data relay backbone~\cite{huangEfficientDifferentiatedRouting2024} rather than intelligent control entities, missing opportunities for distributed control optimization.

\subsection{Controller Placement and Domain Partitioning}

The controller placement problem in satellite networks builds upon terrestrial SDN research but faces unique challenges due to mobility and FOV constraints. Static placement approaches optimize controller locations based on network topology snapshots, achieving good performance under stable conditions but failing to adapt to satellite dynamics.

Dynamic placement strategies~\cite{liStableHierarchicalRouting2024, anushaFeatureSelectionUsing2015} periodically recalculate optimal controller positions and domain assignments. While more adaptive, these approaches often treat mobility as a disturbance rather than leveraging predictable orbital mechanics for stability.

Existing domain partitioning algorithms primarily focus on minimizing communication costs or balancing computational loads, but fail to account for the fundamental constraint that controllers can only manage satellites within their FOV. This oversight leads to control domains that span multiple visibility regions, necessitating costly multi-hop control communications.

\subsection{Mobility-Aware Network Management}

Recent research has begun exploring mobility patterns in satellite networks, but with limited scope. Orbital mechanics integration~\cite{renCircularlyPolarizedSpaceborne2019} primarily focuses on link prediction and handover optimization rather than control domain stability. Movement pattern exploitation~\cite{plastikovHighgainMultibeamBifocal2016} has been applied to traffic routing and resource allocation but not to control plane optimization.

The concept of predictable mobility has been extensively studied in terrestrial mobile networks and vehicular networks, demonstrating significant benefits for proactive resource management. However, satellite networks present unique opportunities due to the completely predictable nature of orbital mechanics, which existing work has not fully exploited for control domain design.

\subsection{Standardization and Research Efforts}

Parallel to academic research on satellite control architectures, the IETF and IRTF have developed foundational protocols and frameworks addressing the dynamic challenges in satellite communication\cite{centenaroSurveyTechnologiesStandards2021}.

The principles for managing highly dynamic topologies, first explored by the IETF in Mobile Ad Hoc Networks (MANET), are highly relevant to controlling ISLs in LEO constellations. Protocols like OLSR demonstrated methods for maintaining routing in the face of constant link changes but also highlighted the substantial control overhead involved. Similarly, the IRTF's research on Delay-Tolerant Networking (DTN), culminating in the Bundle Protocol, directly addresses the long-delay and disruption challenges inherent in MEO/GEO and deep-space links, offering a paradigm for asynchronous communication. While these efforts provide foundational data plane and transport solutions, the organization of a scalable and efficient control plane remains an open challenge.

\subsection{Differentiation from Existing Approaches}

The closest related work is ILA~\cite{huangEfficientDifferentiatedRouting2024}, which also considers MEO-LEO control relationships but focuses primarily on maximizing inter-satellite link utilization rather than minimizing control overhead. AROA~\cite{liOptimizedControllerProvisioning2023} addresses similar load balancing concerns but relies on LEO-based controllers and lacks FOV-aware optimization. Our experimental comparison demonstrates Eunomia's superiority across multiple performance dimensions while providing the first comprehensive evaluation of movement-aware domain partitioning in satellite networks.

\section{Conclusion}\label{sec_con}

This paper investigates the fundamental challenge of control domain partitioning in dynamic hierarchical satellite networks, a key issue in managing emerging mega-constellations. We introduce Eunomia, a novel framework designed to address the intrinsic complexity and high mobility of such networks. Eunomia’s central contribution is a movement-aware partitioning strategy that exploits the predictability of orbital mechanics. By aligning domain boundaries with satellite movement patterns, it establishes stable control domains while effectively reducing control overhead and migration frequency. Through a progressive three-step optimization, Eunomia achieves balanced network load distribution and enhanced algorithmic efficiency. Emulation-based evaluations demonstrate that Eunomia significantly improves both network stability and control efficiency compared to state-of-the-art approaches.

Furthermore, this work opens several promising avenues for future research. Key directions include integrating fault-tolerant domain design to enhance network resilience, developing energy-aware optimization strategies to extend satellite operational lifetimes, and ultimately validating the framework on operational satellite testbeds to bridge the gap between theoretical models and real-world deployment.

%-------------------------------------------------------------------------------
\bibliographystyle{IEEEtran}
\bibliography{ref}

\vspace{-30pt}

% -----作者---------------------------------------------------------------------
\begin{IEEEbiography}[{\includegraphics[width=1in,height=1.25in,clip,keepaspectratio]{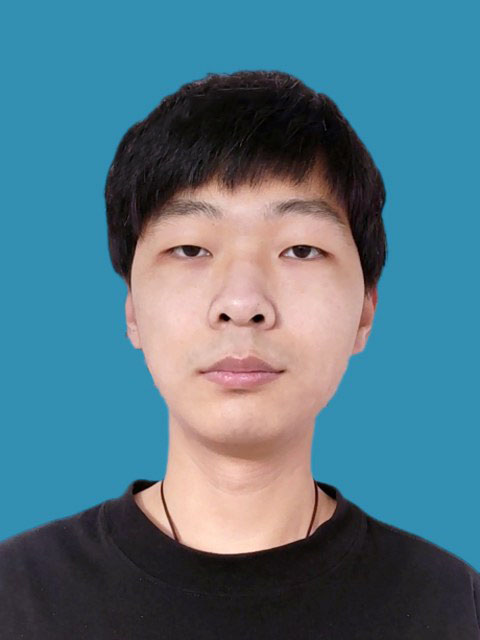}}]{Qi Zhang}
earned his B.Eng. degree from Henan University in 2020 and his M.Eng. in Computer Science and Technology from Soochow University in 2023. He is currently pursuing his PhD at Fudan University. Zhang has made significant contributions to the field of edge computing. His research primarily focuses on routing optimization in integrated space-air-ground networks, a cutting-edge area with vast potential applications.
\end{IEEEbiography}
\vspace{-36pt}

\begin{IEEEbiography}[{\includegraphics[width=1in,height=1.25in,clip,keepaspectratio]{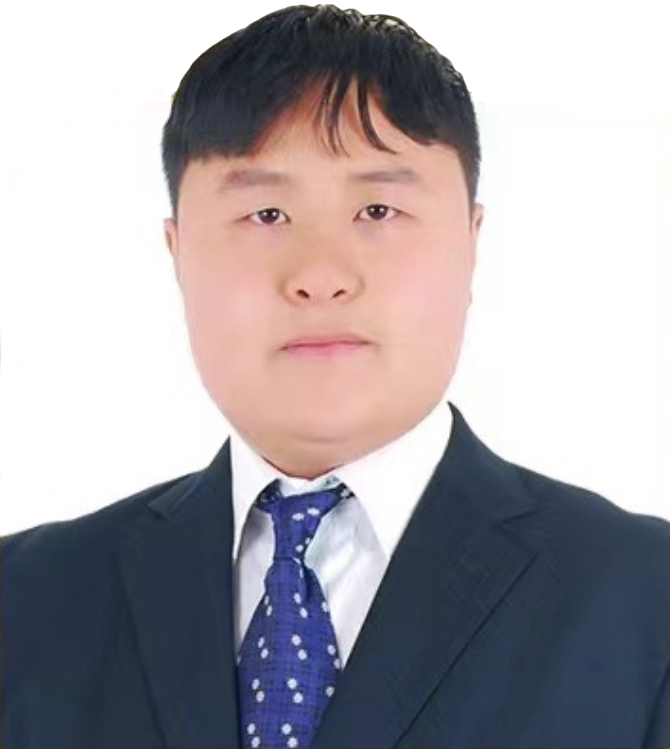}}]{Kun Qiu}
received his B.Sc. from Fudan University in 2013 and his PhD from Fudan University in 2019. He works for Intel as a software engineer from 2019 to 2023. He joined Fudan University in 2023 as an Assistant Professor in the School of Computer Science at Fudan University. His research interests include computer networks and computer architecture. He is a member of IEEE, ACM, and CCF.
\end{IEEEbiography}

\vspace{-36pt}

\begin{IEEEbiography}[{\includegraphics[width=1in,height=1.25in,clip,keepaspectratio]{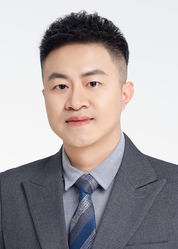}}]{Zhe Chen}
received his PhD in Computer Science from Fudan University, China, with a 2019 ACM SIGCOMM China Doctoral Dissertation Award. He is an Assistant Professor in the School of Computer Science at Fudan University. His research achievements, along with his efforts in launching products based on them, have thus earned him 2021 ACM SIGMOBILE China Rising Star Award recently.
\end{IEEEbiography}

\vspace{-36pt}

\begin{IEEEbiography}[{\includegraphics[width=1in,height=1.25in,clip,keepaspectratio]{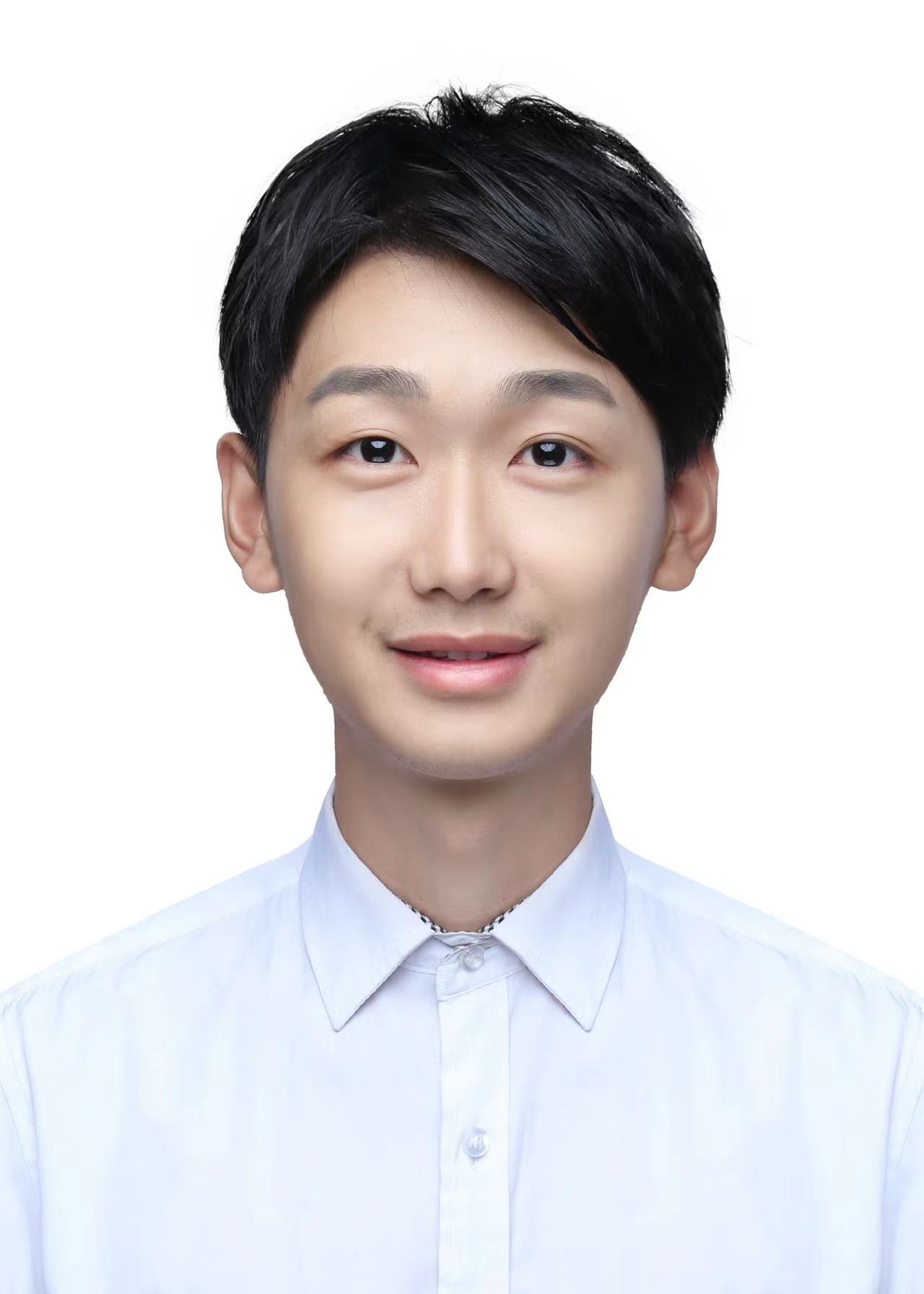}}]{Wenjun Zhu}
received his master's degree from Nanjing University of Posts and Telecommunications in 2020. He works for Intel as a software engineer from 2020 to 2023. Now, he joined Fudan University in 2023 as a software engineer. His research interests include computer networks and computer architecture.
\end{IEEEbiography}

\vspace{-36pt}

\begin{IEEEbiography}[{\includegraphics[width=1in,height=1.25in,clip,keepaspectratio]{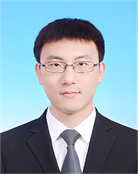}}]{Xiaofan Xu}
received his bachelor's degree from Peking University and his doctoral degree from University of Heidelberg. He is an expert in satellite communication technologies. He is currently the director of the State Key Laboratory of Satellite Network and the Shanghai Key Laboratory of Satellite Network. His research interests include 5G/6G communication technology and SAGIN architecture.
\end{IEEEbiography}

\vspace{-36pt}

\begin{IEEEbiography}[{\includegraphics[width=1in,height=1.25in,clip,keepaspectratio]{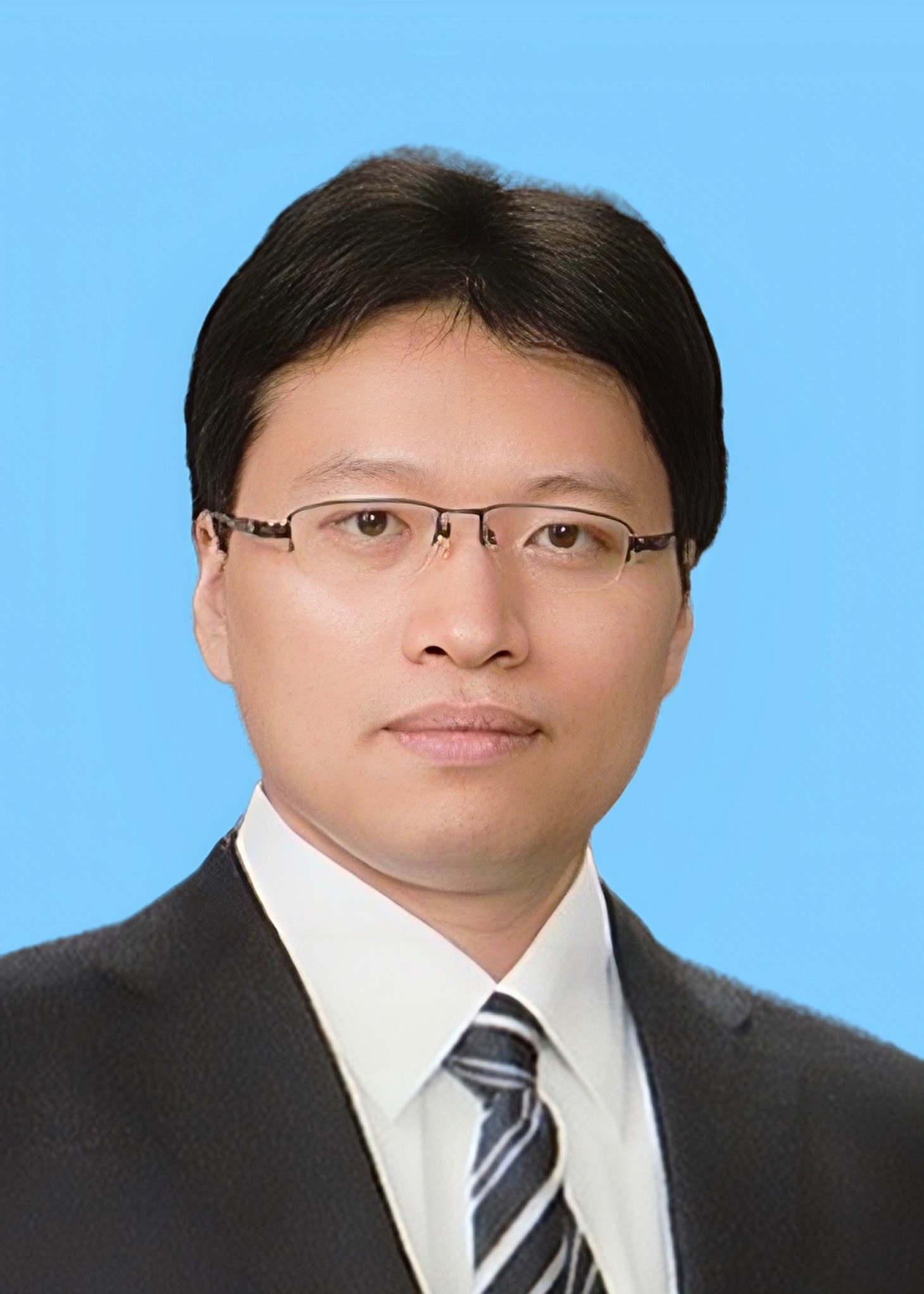}}]{Ping Du}
is a senior expert in the State Key Laboratory of Satellite Network and the Shanghai Key Laboratory of Satellite Network. His research interest includes architecture of satellite network, mobile communication etc.
\end{IEEEbiography}

\vspace{-36pt}

\begin{IEEEbiography}[{\includegraphics[width=1in,height=1.25in,clip,keepaspectratio]{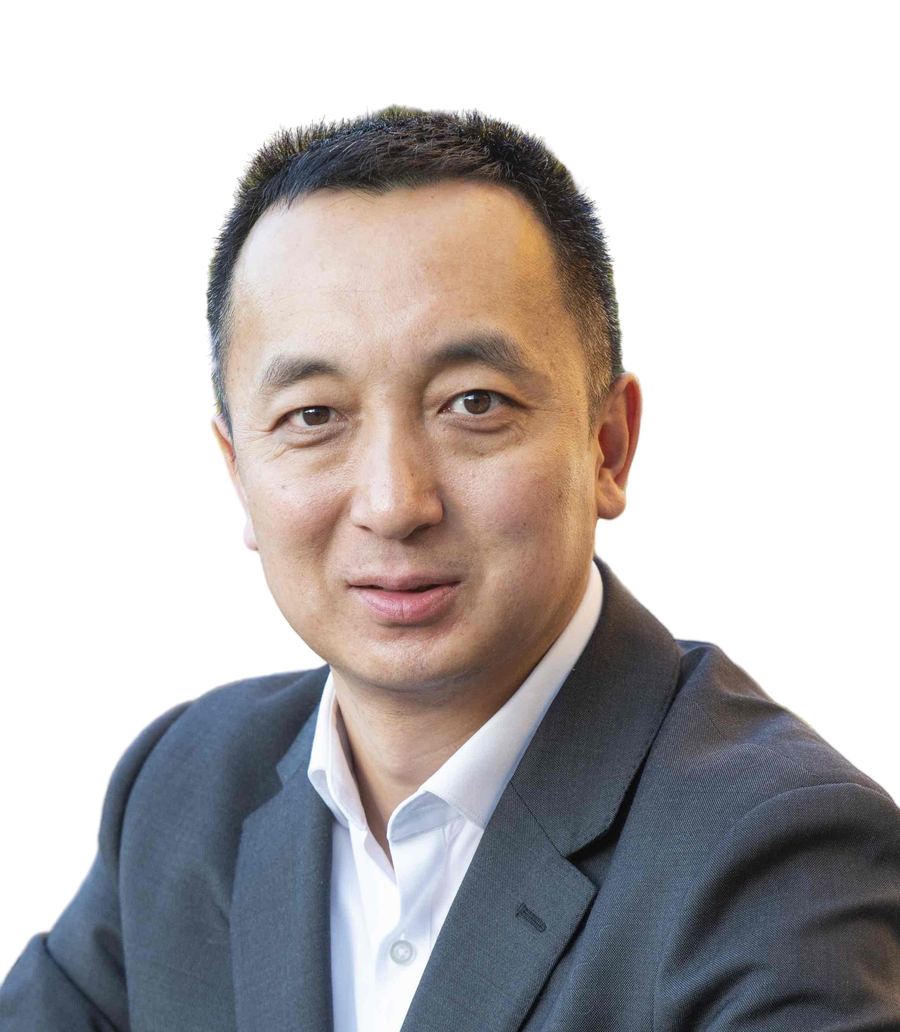}}]{Yue Gao}
received his PhD from the Queen Mary University of London, UK, in 2007. He is a Chair Professor at the School of Computer Science, Director of the Intelligent Networking and Computing Research Centre at Fudan University, China and a Visiting Professor at the University of Surrey, UK. His research interests include smart antennas, sparse signal processing and cognitive networks for mobile and satellite systems. He is a Fellow of the IEEE.
\end{IEEEbiography}

%%%%%%%%%%%%%%%%%%%%%%%%%%%%%%%%%%%%%%%%%%%%%%%%%%%%%%%%%%%%%%%%%%%%%%%%%%%%%%%%
\end{document}